\documentclass[twocolumn,aps,showpacs,prl,amsmath,amssymb,floatfix,superscriptaddress]{revtex4-1}
\usepackage{color}
\usepackage{graphicx}% Include figure files
\usepackage{dcolumn}% Align table columns on decimal point
\usepackage{bm}% bold math
\usepackage{array}
\usepackage{float}
\usepackage{supertabular}
\usepackage{longtable}
\usepackage{mathrsfs}
\usepackage{txfonts}
\usepackage{wasysym}
\usepackage{hyperref}

\newcommand{\beginsupplement}{%
	\setcounter{table}{0}
	\renewcommand{\thetable}{S\arabic{table}}%
	\setcounter{figure}{0}
	\renewcommand{\thefigure}{S\arabic{figure}}%
	\setcounter{equation}{0}
	\renewcommand{\theequation}{S\arabic{equation}}%
	\setcounter{section}{1}
	\renewcommand{\thesection}{S\arabic{section}}%
}

\begin{document}

\title{Optimal Control of Majorana Zero Modes}

\author{Torsten Karzig}
\affiliation{
Institute of Quantum Information and Matter, California Institute of Technology, Pasadena, California 91125, USA} 

\author{Armin Rahmani}
\affiliation{
Theoretical Division, T-4 and CNLS, Los Alamos National Laboratory, Los Alamos, New Mexico 87545, USA} 

\affiliation{Quantum Matter Institute, University of British Columbia, Vancouver, British Columbia, Canada V6T 1Z4}

\author{Felix von Oppen}
\affiliation{
Dahlem Center for Complex Quantum Systems and Fachbereich Physik, Freie Universit\"at Berlin, 14195 Berlin, Germany}

\author{Gil Refael}
\affiliation{
Institute of Quantum Information and Matter, California Institute of Technology, Pasadena, California 91125, USA} 

\date{\today}

\begin{abstract}
Braiding of Majorana zero modes provides a promising platform for quantum information processing, which is topologically protected against errors. Strictly speaking, however, the scheme relies on infinite braiding times as it utilizes the adiabatic limit. Here we show how to minimize nonadiabatic errors for finite braiding times by finding an optimal protocol for the Majorana movement. Interestingly, these protocols are characterized by sharp transitions between Majorana motion at maximal and minimal velocities. We find that these so-called bang-bang protocols can minimize the nonadiabatic transitions of the system by orders of magnitude in comparison with naive protocols.
\end{abstract}
 
\maketitle

Topological quantum computing is a promising approach to quantum information processing, which provides remarkable robustness against errors~\cite{kitaev_fault-tolerant_2003,nayak_non-abelian_2008}. At the heart of this approach lie exotic quasiparticles known as non-Abelian anyons, which can emerge in several condensed matter systems; \textit{adiabatic} exchange of such quasiparticles transforms the many-body wave function to a different degenerate wave function, in turn processing the information stored (nonlocally) in these quantum wave functions. In fact, adiabatic exchange, i.e., braiding, is the key ingredient of topological quantum computing. However, perfect adiabaticity requires infinite times. Therefore, it is imperative to be able to perform such transformations in finite time, while minimizing the undesirable nonadiabatic effects~\cite{cheng_nonadiabatic_2011,perfetto_dynamical_2013,karzig_boosting_2013,scheurer_nonadiabatic_2013,amorim_majorana_2014}.

Majorana zero modes are one of the simplest and most important non-Abelian quasiparticles \cite{alicea_new_2012,beenakker_search_2013}. There have been several proposals~\cite{fu_superconducting_2008,lutchyn_majorana_2010,oreg_helical_2010,duckheim_andreev_2011,nadj-perge_proposal_2013}, as well as experimental progress~\cite{mourik_signatures_2012,das_zero-bias_2012,deng_anomalous_2012,rokhinson_fractional_2012,finck_anomalous_2013,pribiag_edge-mode_2014,hart_induced_2014,nadj-perge_observation_2014}, toward realizing these modes in one-dimensional hybrid systems, e.g., semiconducting quantum wires coupled to superconductors. Making a network of such quantum wires can in turn allow for braiding of these Majorana modes~\cite{alicea_non-abelian_2011}. Thus, the minimal building block of quantum information processing with the quantum-wire incarnation of Majorana zero modes is moving them along the wire adiabatically. These zero modes are bound to domain walls between the topological and nontopological phases, whose position and velocity can be tuned externally, e.g., by means of gate electrodes. Adiabatic transport of the Majoranas then amounts to slowly moving these domain walls.

\begin{figure}[center]
	\includegraphics[width=7.5cm]{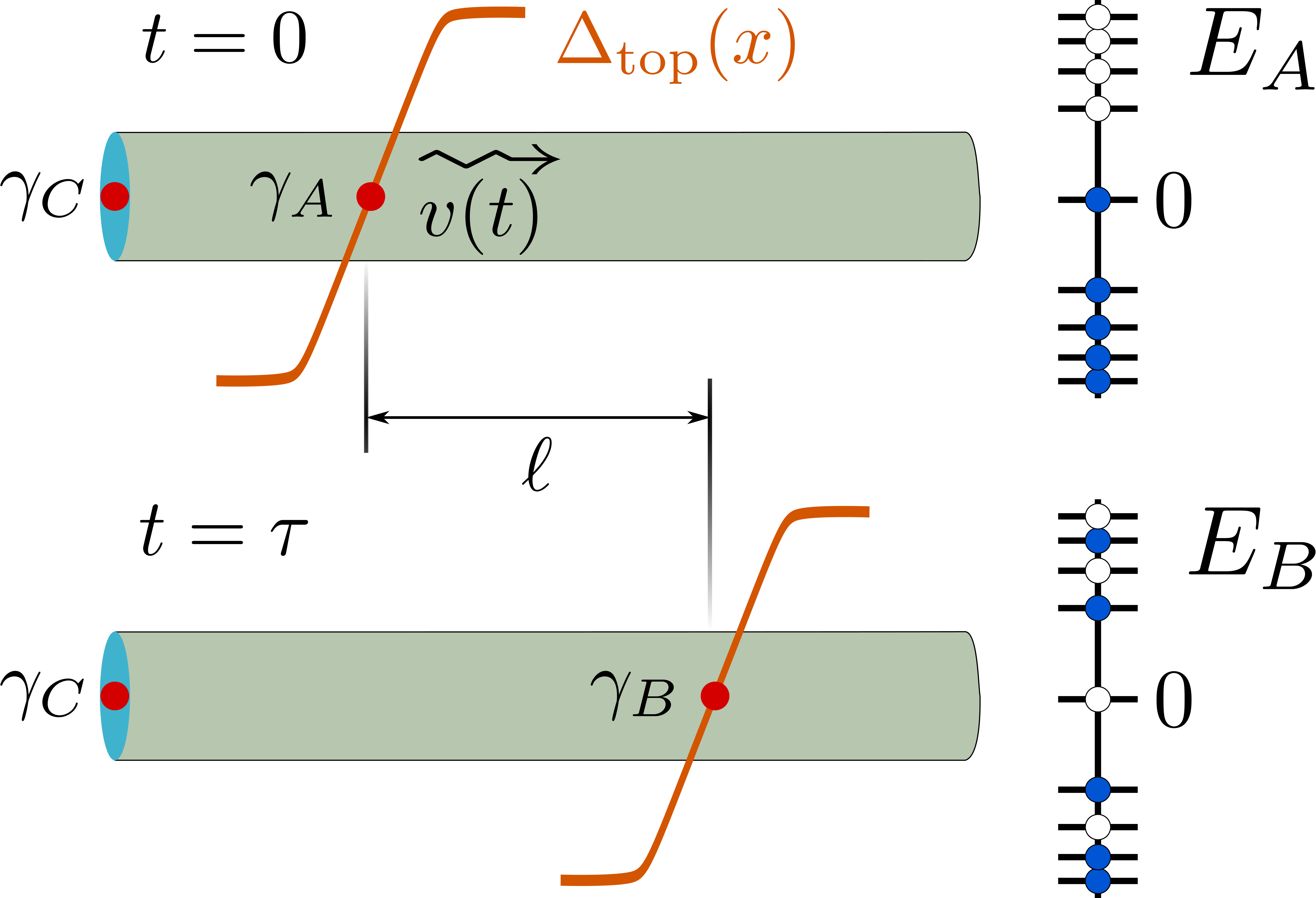}
	\caption{Nonadiabatic motion of Majorana bound states. When moving a Majorana-carrying domain wall in a finite time $\tau$ by a distance $\ell$, the final state will in general experience nonadiabatic excitations as indicated by the difference in the occupation of the low energy bound states, before (upper panel, $t=0$, position $A$), and after (lower panel, $t=\tau$, position $B$) the motion. }
	\label{fig:Setup}
\end{figure}

Consider a Majorana mode in a quantum wire bound to a domain wall at point $A$, with the system in one of its ground states (Fig.~\ref{fig:Setup}) and imagine moving the domain wall (and hence the associated Majorana mode) to point $B$ a distance $\ell$ away within a prescribed time $\tau$. What is the optimal choice for the time-dependent velocity of the domain wall? As this translation is carried out in finite time, there are deviations from the fully adiabatic evolution. We would like to choose a protocol which generates a state as close as possible to adiabatically moving the domain wall to point $B$. This is clearly  important for realizations of topological quantum computers as both practical performance considerations and parasitic decoherence processes such as quasiparticle poisoning limit the available time for braiding processes~\cite{goldstein_decay_2011,budich_failure_2012,rainis_majorana_2012,mazza_robustness_2013,ng_decoherence_2014}.

More broadly, optimal control has emerged as a new direction in quantum dynamics~\cite{hohenester_optimal_2007,salamon_maximum_2009,chen_fast_2010,rahmani_optimal_2011,doria_optimal_2011,caneva_speeding_2011,choi_optimized_2011,del_campo_fast_2011,hoffmann_time-optimal_2011,rahmani_cooling_2013}. By finding the \textit{best} protocols to optimize a certain figure of merit, quantum optimal control paves the way towards harnessing the power of quantum evolution. While the primary motivation for the field comes from experimental advances with ultracold atoms, the applicability of quantum optimal control goes well beyond these systems. The subject of this paper, i.e., finding the optimal protocol to move a Majorana mode along a quantum wire, shows that optimal control can play an important role in topological quantum computing. 

{\em Figure of merit.---}We start by defining an appropriate figure of merit. A very natural choice in the present case is to minimize
\begin{equation}
	c(\tau)=1-|\langle\Psi_{B}^{\rm ad}|\Psi(\tau)\rangle|^2,
	\label{eq:costfunction}
\end{equation}
which quantifies the deviations from the adiabatic evolution in terms of the squared overlap between $|\Psi(\tau)\rangle=U(\tau)|\Psi(0)\rangle$, the wave function of the system obtained after the quantum evolution for a time $\tau$ [with evolution operator $U(\tau)$], and $|\Psi^{\rm ad}_{B}\rangle$, the wave function after a perfectly adiabatic evolution \footnote{Note that in the context of Majorana quenches a similar figure of merit was recently studied in Ref.~\cite{hegde_quench_2014}}. In the present case, $|\Psi^{\rm ad}_{B}\rangle$ is simply the ground state of the Hamiltonian with the domain wall at position $B$, while the initial state $|\Psi(0)\rangle$ is the ground state with the domain wall at point $A$. In general, the above cost function is vulnerable to the orthogonality catastrophe for infinite systems. Here, however, we restrict our Hilbert space to the discrete bound states within the (bulk) gap to the continuum.

Strictly speaking, the topological protection is lost if the system strays too far from the instantaneous ground state and the Majorana mode leaks to the continuum (separated by the bulk gap). Here we consider permissible velocities $v(t)<v_{\rm max}$ so that we are never too far from the adiabatic limit with respect to the bulk gap. The evolution, however, does create nonadiabatic excitations within the bound-state spectrum of the domain wall, which are corrected by our optimization scheme.
 
\begin{figure}[center]
	\includegraphics[scale=1]{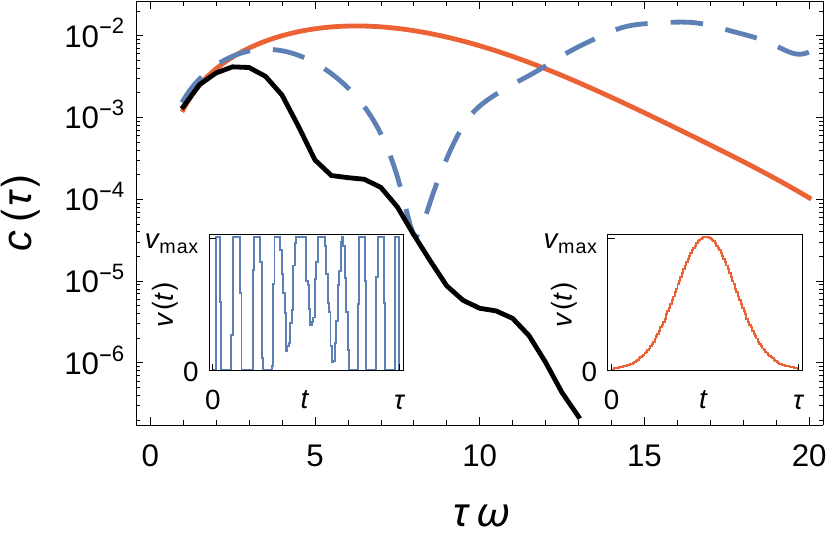}
	\caption{
Cost function for bang-bang-type optimal protocols (Gaussian reference protocol) shown in black (red) as a function of $\tau$ (with fixed average and maximum velocity). Optimal protocols were individually obtained for each $\tau$ (left inset shows a protocol optimized for $\tau=8/\omega$), while the reference protocol is a smooth Gaussian curve shown in the right inset. The cost function $c(\tau)$ can be reduced by several orders of magnitude when using optimal protocols. The dashed blue curve shows the cost function obtained by applying the optimal protocol shape corresponding to $\tau=8/\omega$ to other times. It outperforms the reference protocol for a wide time interval. Numerical parameters used: $v_{\rm max}=0.3u$, $N=128$, $n_{\rm max}=30$, $n_{c}=7$.}
	\label{fig:costfunction}
\end{figure}

We use Monte Carlo calculations (simulated annealing) to find the optimal protocol which minimizes the cost function in Eq.~(\ref{eq:costfunction}) for a fixed total time $\tau$, average velocity $\ell/\tau$, and maximal velocity $v_{\rm max}$. This method finds the optimal protocol without making any \textit{a priori} assumptions. Remarkably, we find that the optimal protocols have a bang-bang form, i.e., they are a sequence of sudden quenches between the maximal ($v_{\rm max}$) and the minimal (0) allowed velocities. Despite ubiquitously occurring in optimal control theory \cite{salamon_maximum_2009}, such bang-bang protocols appear quite counterintuitive in the present context. Nevertheless, we find that they reduce the nonadiabatic errors by orders of magnitude in comparison with simple nonoptimal protocols, which one may construct intuitively (see Figure~\ref{fig:costfunction}). In addition to our numerical results, which are obtained for specific models of the domain wall, we also adapt Pontryagin's maximum principle to our problem and establish more generally that the optimal protocols must be bang-bang.
 
\emph{Model.---}We consider the effective Hamiltonian for a quantum wire (or topological insulator edge) \cite{fu_superconducting_2008,lutchyn_majorana_2010,oreg_helical_2010} in the vicinity of a topological domain wall, assuming that the gap varies linearly as a function of position~\cite{karzig_boosting_2013}
\begin{equation}
  \hat{H}=\int \hat{\Psi}^\dagger(x) {\cal H}\hat{\Psi}(x)d x , \quad {\cal H}=-iu\partial_x\sigma_z-b(x-y)\sigma_x.
  \label{eq:Hamiltonian}
\end{equation}
Here, $\sigma_{i}$ are Pauli matrices and $\hat{\Psi}^\dagger(x)=\left(\hat{\psi}^\dagger_{\uparrow}(x)+\hat{\psi}_\uparrow(x),\hat{\psi}^\dagger_{\downarrow}(x)-\hat{\psi}_\downarrow(x)\right)$ with $\hat{\psi}_\uparrow(x)$ [$\hat{\psi}_\downarrow(x)$] representing the fermionic annihilation operator of spin up (down) electrons at position $x$. The parameter $y$ denotes the position of the domain wall and is time dependent when the domain wall is moving along the wire. 

For fixed $y$, the above Hamiltonian gives rise to single-particle bound states $\hat{\gamma}_{n,y}$ localized at $x=y$ with the spectrum $\varepsilon_n={\rm sign}(n)\sqrt{|n|}\omega$, where $n$ runs from $-\infty$ to $\infty$. The corresponding wave functions $\phi_n=(\mathrm{i} +\sigma_x)(\mathrm{sign}(n)g_{|n|-1},g_{|n|})/2$ are given in terms of harmonic oscillator eigenstates $g_n(x-y)$ with frequency $\omega=\sqrt{2ub}$ and oscillator length $\xi=\sqrt{u/b}$. It can be shown that the zero-energy state $\phi_0$ is a Majorana state with quasiparticle operator  $\hat{\gamma}_{0,y}=\hat{\gamma}^\dagger_{0,y}$~\cite{karzig_boosting_2013}.
We assume that the domain wall is initially at $y(0)=0$ (point $A$). The velocity $v(t)={d\over dt } y(t)$ of the domain wall is then subject to the following constraints: $0\leqslant v(t)\leqslant v_{\rm max}$ and $y(\tau)=\int_0^\tau dt v(t)=\ell$. To avoid the superluminal regime, where the bound states become unstable \cite{karzig_boosting_2013}, we work at velocities $v_{\rm max} < u$.

Physically, the linear form  $b(x-y)$  of the domain wall extends over a finite length scale. this implies that we have a finite number of bound states and then a continuum of excitations. We implement this by using two cutoffs: the time evolution is done within the bound state spectrum with $|n|<n_{\rm max}$, where very large $n$ model the continuum. The cost function is computed by projecting the wave functions onto a smaller Hilbert space with $|n|<n_c$. Physically, $n_c$ represents the number of bound states. We can relate the cost function to occupation numbers $\hat{n}_i$ with $\hat{n}_{i\neq0}=\hat{\gamma}_{i,B}^\dagger\hat{\gamma}_{i,B}$. The Majorana mode requires special treatment. We define the delocalized fermionic zero mode $\hat{d}_{0}=(\hat{\gamma}_{0,B}+{\rm i}\hat{\gamma}_{0,C})/\sqrt{2}$ (note that $\hat{\gamma}_{0,C}$ is static). Then, we write the corresponding occupation number $\hat{n}_{0_-}=\hat{d}_0^\dagger\hat{d}_0$, assuming $\langle\Psi_B^{\mathrm{ad}}|\hat{n}_{0_-}|\Psi_B^{\mathrm{ad}}\rangle=1$ without loss of generality. The minus (plus) subscript indicates that $\hat{n}_{0_-}$ ($\hat{n}_{0_+}=\hat{d}_0\hat{d}_0^\dagger$) should be treated like the other negative-energy (positive-energy) states.

For small maximal velocities $v_{\rm max}$, the occupation numbers $\hat{n}_{i\leq0}$ are still close to unity, which allows for an expansion of $\hat{n}_{-i}=1-\hat{n}_{i}$  in small $\hat{n}_{i}$ (with $i\geq0_+$).  The cost function can then be approximated as
\begin{equation}
	c(\tau)\approx\sum_{n_c>i\geq 0_+} \langle\hat{n}_i\rangle_{\tau}-\sum_{n_c>j>i,\,n_c>i\geq 0_+} \langle\hat{n}_i\hat{n}_j\rangle_{\tau}\,,
	\label{eq:costfunction2}
\end{equation}
which may be evaluated straightforwardly in the Heisenberg picture by computing  operators $\hat{\gamma}_{n,B}(\tau)$ (and $\hat{d}_{0}(\tau)$). We have made use of the fact that the cost function is an expectation value of the Heisenberg evolved projector $|\Psi_{B}^{\mathrm{ad}}\rangle\langle\Psi_{B}^{\mathrm{ad}}|=\Pi_{ i\leq0}\hat{n}_i$.

We evaluate the Heisenberg operators by approximating the protocol for moving the domain wall by a  piece-wise constant sequence of velocities $v_i$ (each of duration $\delta t$) for $i=1\dots N$. For each piece, the time evolution can be described by a mapping to the static case by a Lorentz boost, with boosted bound-state wavefunctions $\phi_n^{(v_i)}(x-v_i t)$ and a renormalized spectrum $\varepsilon_n^{(v_i)}$ \cite{karzig_boosting_2013} (see also Supplemental Information). With these exact constant-velocity solutions, the Heisenberg evolution of the domain wall bound states takes the form
\begin{equation}
	U(\tau)^\dagger \hat{\gamma}_{n,B} U(\tau) = \sum_{\{m_i\}}a^{(v_N)}_{n,m_N}\,\dots\,a^{(v_1)}_{m_2,m_1} \hat{\gamma}_{m_1,A}, 
	\label{eq:Heisenberg}
\end{equation}
where $U(\tau)$ is the full many body time evolution operator and  $a^{(v)}_{n,m}=\sum_k\langle\phi^{(0)}_n|\phi^{(v)}_k\rangle\langle\phi^{(v)}_k|\phi^{(0)}_m\rangle \exp\big(-{\rm i}\varepsilon_k^{(v)}\delta t\big)$. The matrix elements $\langle\phi^{(0)}_n|\phi^{(v)}_k\rangle$ are essentially overlaps of harmonic oscillator wavefunctions shifted by $\sim\!\!\sqrt{k}v/u\xi$ relative to each other. For small velocities, we have $\langle\phi^{(0)}_n|\phi^{(v)}_k\rangle \propto (v/u)^{(|n|-|k|)}$ \footnote{With the exception {$\langle\phi^{(0)}_n|\phi^{(v)}_{-n}\rangle \propto (v/u)^{2}$}. See also Supplemental Information.}. The sums over the states (denoted by the indices $k$ and $m_i$) can thus be cut off at a large $n_{\rm max}$ for numerical evaluation.

\emph{Optimization.---}Based on the cost function \eqref{eq:costfunction2}, we use simulated annealing to find the optimal protocol \cite{rahmani_optimal_2011,rahmani_quantum_2013}. In this method, we fix the total time $\tau$ and distance $\ell$ and use a piecewise-constant protocol with $N$ pieces of duration $\delta t=\tau/N$.  (We then increase $N$ systematically until convergence). We implement the constraint of a fixed average velocity in each Monte Carlo step by increasing the velocity of one randomly chosen interval while decreasing the velocity of another by the same amount. If the change $\Delta c$ in the cost function is negative, we accept the move. Otherwise, we accept it with probability $e^{-\Delta c/T_{MC}}$, where $T_{\rm MC}$ is a fictitious temperature that is  gradually reduced to zero.

 %(Note that our scheme also works in the opposite limit where relaxation is absent such that high energy excitations decouple from the low energy bound states.) 

As mentioned above, we only include $n_{\mathrm{max}}$ bound states in the numerical optimization. This makes the time evolution of states close to $n_{\rm max}$ unreliable. Since the cost function is evaluated using a smaller cutoff  $n_c\ll n_{\rm max}$, corresponding to the physical number of bound states, our results are independent of $n_{\rm max}$. Note that the optimization is aimed at conserving the overall parity of the bound states, which ultimately protects the Majorana qubit \cite{akhmerov_topological_2010}. The states $|n|>n_c$ that are left out from the optimization would represent high-energy continuum states, with nonadiabatic occupations that are not necessarily weaker for the optimal protocol than for a naive protocol. They are, however, naturally suppressed if the protocols are slow with respect ot the inverse bulk gap. Moreover, excitations in these states do not affect the parity of the delocalized fermionic mode, i.e.,  $i\gamma_B\gamma_C$ (see Supplemental Material for details).
\begin{figure}[center]
	\includegraphics[scale=1.25]{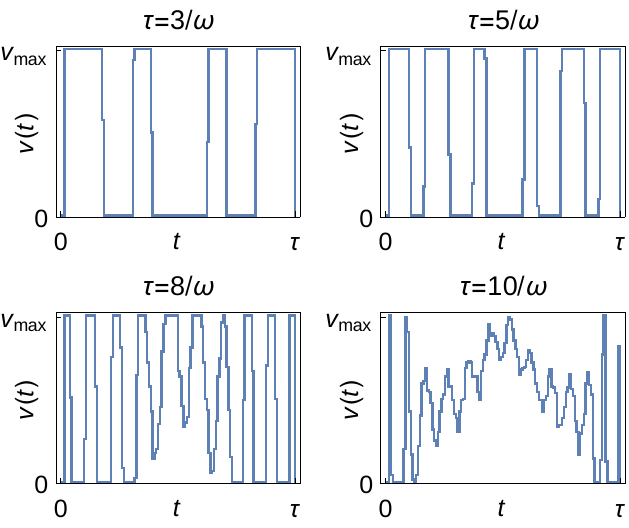}
	\caption{Optimal bang-bang-type protocols for different durations $\tau$. The number of bangs increases with $\tau$. Due to the finite number of time steps, here $N=128$, this leads to numerical artifacts for large times where the size of the bangs reaches the time step width $\delta t$. The optimal protocols are then smoothed out because of an effective averaging over times $\delta t$. Further numerical parameters used: $v_{\rm max}=0.3u$, $n_{\rm max}=30$, $n_{c}=7$.}
	\label{fig:bang-bangs}
\end{figure}

\begin{figure}[center]
	\includegraphics[scale=1]{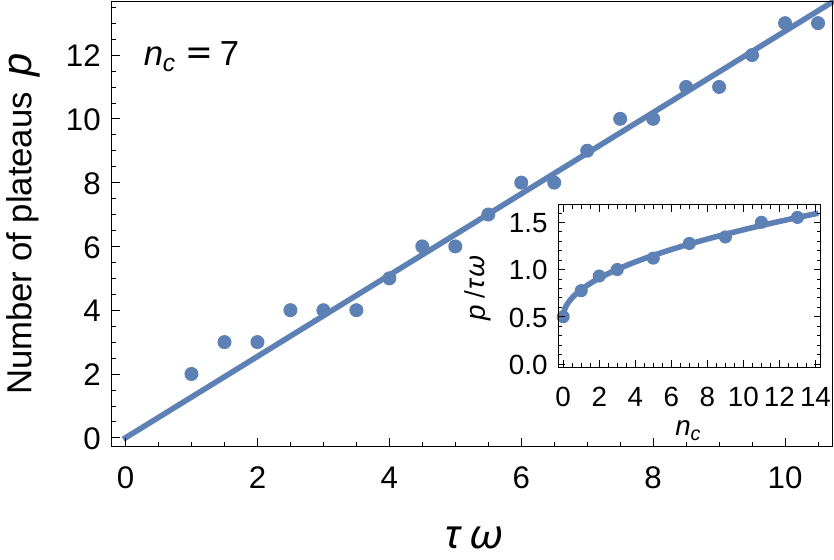}
	\caption{Dependence of the bang-bang-type protocols on the duration $\tau$. The number of plateaus with high velocity $p$ scales linearly to the protocol duration $\tau$. The inset shows the change of the slope $p/\tau\omega$ with the number of bound states in the cost function $n_c$. A fit to our data shows that it can be well approximated by $p/\tau\omega=0.3\sqrt{n_c}+0.5$\,.}
	\label{fig:number of bangs}
\end{figure}

\emph{Results.---}The central result of our Monte Carlo simulations is that the optimal protocols are of bang-bang character and outperform naive protocols by orders of magnitude (see Fig.~\ref{fig:costfunction}). The sharp bang-bang transitions can be very well resolved numerically for not-too-large $\tau$ (see Fig.~\ref{fig:bang-bangs}). For a fixed number of velocity steps $N$, the time resolution decreases for larger $\tau$. Once the minimal time steps $\delta t=\tau/N$ exceed the interval between consecutive velocity jumps of the optimal protocols, the numerics average the optimal protocol over times $\delta t$ resulting in a smoothing of the bang-bang character. Thus, when taking the adiabatic limit $\tau \rightarrow \infty$ before increasing $N\rightarrow\infty$, the optimal protocols become smooth and are determined by the density of underlying high-velocity sections. 

For good time resolutions, the main characteristic of the optimal protocols is the number $p$ of high-velocity plateaus. Interestingly, $p$ is independent of many of the specifics of the braiding process such as the maximal velocity $v_{\max}$ and the displacement $\ell$ (controlling the average velocity for fixed $\tau$), which only change the size of the plateaus. Instead, we find that $p$ is determined by the domain-wall spectrum. More specifically, $p/\tau$ is of the order of the bound state energy (see Fig.~\ref{fig:number of bangs}). In a simple picture, the bang-bang protocols can be thought of as well timed echos that reverse the nonadiabatic evolution. From this point of view, it is natural to assume that the relevant scale for this timing is given by the energy of the excited bound states. This is in line with the approximate $p/\tau\sim \omega \sqrt{n_c}=\varepsilon_{n_c}$ behavior that we observe in our simulations (see Fig.~\ref{fig:number of bangs}). We will see below that $\omega\sqrt{n_c}$ also appears as a characteristic frequency scale of the corresponding Pontryagin equations that describe the optimal protocol.

Although the form of the protocols does not converge for large $n_{c}$, the changes become less and less important for the cost function. Our data indicate (see Supplemental Information) that $c(\tau)$ saturates for large $n_{c}$. Similarly, when applying the cost function (with fixed  $n_{c}$) to protocols optimized for ``wrong" values of $n_{c}'$, their performance is still very close to the $n_c'=n_c$ case for not too small values of $n_c',n_c$. Specifically, even though a large $n_c$ yields an optimal protocol with large $p$, it can still be well approximated by a protocol with fewer bangs that would be obtained when choosing a smaller $n_c'$. All these observations reflect the weak occupation of states with large quantum numbers due to the weakly diabatic. 

\emph{Pontryagin equation.---}{We now prove that the optimal protocols must be bang-bang by using a generalization of the calculus of variations known as Potryagin's principle \cite{pontryagin_mathematical_1987}. We briefly review the formalism. Assume we have a set of dynamical variable $X(t)$ that evolve with the equations of motion
$\dot{{X}}_j=f_j(\{{X}\},v)$, boundary conditions ${X}_j(0)={X}_j^0$, and permissible control $v(t)$. (In our case, these variables correspond to some parameterization of the wave function.) For a given control the equations of motion then determine the the dynamical variables as a function of time. 

We would like to find the optimal control $v^*(t)$ that minimizes a general cost function $c(\{{X(\tau)}\})+\int_0^\tau {\cal L}(\{X\},v)dt$. The function $c(\{{X(\tau)}\})$ only depends on the final values of the dynamical variables at $t=\tau$, while the cost additional integral over ${\cal L}(\{X\},v)$ allows for dependence on the entire trajectory. We can think of the equations of motion above as constraints that can be implemented by Lagrange multipliers $P_j(t)$ (hereafter referred to as conjugate momenta) at every point in time by considering a constfunction $S=c(\{{X(\tau)}\})+\int_0^\tau dt {\cal L}(\{X\},v)+\sum_j \int_0^\tau dt P_j\left[f_j(\{X\},v)-\dot{X}_j\right]$. Minimizing $S$ (i.e., setting $\delta S=0$) then implies the following equations of motion for the conjugate momenta: $\dot{P}_j=-\frac{\partial {\mathscr H}}{\partial{X_j}}$, with boundary condition $P_j(\tau)={\partial \over  \partial X_j} c(\{X(\tau)\})$,  where the \textit{optimal-control Hamiltonian} is constructed as
\begin{equation}\label{eq:och}
{\mathscr H}(\{X,P\},v)={\cal L}(\{X\},v)+\sum_j  P_j\:f_j(\{X\},v).
\end{equation}
Furthermore, the optimal control $v^*(t)$ and the corresponding $\{X^*,P^*\}$ satisfy 
${\mathscr H}(\{X^*,P^*\},v^*)=\min_{\{v\}}{\mathscr H}(\{X^*,P^*\},v)$. In other words, if we know the optimal trajectories $X^*$ and $P^*$, then at every point in time $v^*$ is a permissible $v$ that minimizes ${\mathscr H}$. An important consequence of this is that if ${\mathscr H}$ is linear in $v$, then depending on the sign of the coefficient of $v(t)$ at any given time (which depends on $P^*(t)$ and $X^*(t)$), \textit{$v(t)$ takes either its minimum or its maximum allowed value, resulting in a bang-bang protocol.}

%Assume we want to find the optimal control $v^*(t)$ that minimizes a cost function $c(\{{x(\tau)}\})+\int_0^\tau {\cal L}(\{x\},v)dt$ for a set of dynamical variables $\{x(t)\}$ evolving with the equations of motion
%$\dot{{x}}_j=f_j(\{{x}\},v)$, boundary conditions ${x}_j(0)={x}_j^0$, and permissible control $v(t)$.  We can think of the equations of motion above as constraints that can be implemented by Lagrange multipliers $p_j(t)$ (hereafter referred to as conjugate momenta) at every point in time by adding $\sum_j \int_0^\tau dt p_j\left[f_j(\{x\},v)-\dot{x}_j\right]$ to the cost function. Calculus of variations then implies the following equations of motion for the conjugate momenta: $\dot{p}_j=-\frac{\partial {\mathscr H}}{\partial{x_j}}$, with boundary condition $p_j(\tau)={\partial \over  \partial x_j} c(\{x(\tau)\})$, where the \textit{optimal-control Hamiltonian} is constructed as
%\begin{equation}\label{eq:och}
%{\mathscr H}(\{x,p\},v)={\cal L}(\{x\},v)+\sum_j  p_j\:f_j(\{x\},v).
%\end{equation}
%The optimal control $v^*(t)$ and the corresponding $\{x^*,p^*\}$ then satisfy 
%${\mathscr H}(\{x^*,p^*\},v^*)=\min_{\{v\}}{\mathscr H}(\{x^*,p^*\},v)$.

In the present case, we have a very similar problem: The control parameter is the velocity $v(t)$ of the domain wall for $0<t<\tau$  and the dynamical variables constitute a parameterization of the time-dependent wave function of the system. Our physical cost function $c(\tau)$ only depends on the final values of the dynamical variables (no dependence on trajectory). However, we have one additional constraint, namely, a fixed total displacement $\ell$, which can be accounted for by adding a Lagrange-multiplier term $\lambda \left(\int_0^\tau v(t)dt -\ell\right)$ to the cost function. This constraint only adds a \textit{linear term} in $v$ to the optimal-control Hamiltonian, i.e., ${\cal L}(\{X\},v)=\lambda v$ [see Eq.~\eqref{eq:och}]. Now we only need to identify a set of dynamical variables with linear equations of motion in $v$ to prove the bang-bang nature of the protocols.

It is convenient to expand the time evolution of the (two-component) single-particle wave functions as 
\begin{equation}
\psi^m(x,t)=\sum_n \left(
\varphi_n^m(t),\:
\theta_n^m(t)
 \right)g_n\left(x-y(t)\right),
 \end{equation}
where $m$ denotes the bound state number of the initial condition $\psi^m(x,0)=\phi_m(x)$.
The shift of $y(t)$ to the instantaneous position of the domain wall allows us to readily relate the dynamical variables, i.e., the real and imaginary parts of $\varphi_n^m$ and $\theta_n^m$, to $|\Psi_B^{\rm ad}\rangle$. The cost function $c(\tau)$ [see Eq.~\eqref{eq:costfunction}] can therefore be obtained from the final values of these dynamical variables. Notice that the harmonic-oscillator eigenstates $g_n(x-y(t))$ provide an orthonormal basis and the dynamical variables are some coefficients. 
As shown in the supplemental material , the equations of motion for these dynamical variables indeed turn out to be\textit{ linear} in $v$, completing the proof for the bang-bang nature of the optimal protocol: 
\begin{equation}\label{eq:eomphi}
\begin{split}
\dot{\varphi}^m_n =
{\omega\over2} (v/u-1)&\left(\sqrt{n+1} \varphi^m_{n+1}-\sqrt{n}\varphi^m_{n-1}\right)\\
&
+i{\omega\over 2}\left(\sqrt{n+1} \theta^m_{n+1}+\sqrt{n}\theta^m_{n-1}\right),
\end{split}
\end{equation}
 and a similar expression with $v/u-1\rightarrow v/u+1$ and $\varphi\leftrightarrow\theta$ for $\dot{\theta}^m_n$.}

%  Therefore, the optimal-control Hamiltonian ${\mathscr H}$ is linear in $v$ and the condition given below Eq.~\eqref{eq:och} for $v^*$ can only be satisfied if $v(t)$ takes its minimum or maximum permissible value as determined by the sign of the coefficient of $v$ [i.e., ${\rm sgn}\left(\partial_v {\mathscr H}\right)$] unless $\partial_v {\mathscr H}=0$ over finite intervals. 

Recall that since the optimal protocol is determined by the sign of  $\partial_v {\mathscr H}$, the discontinuities in the optimal protocols should coincide with zeros of $\partial_v {\mathscr H}$. We have checked this explicitly for our optimal protocols (see the Supplemental Material). Also notice that the distance between these zeros (typical duration of a bang) is determined by the oscillations of ${\mathscr H}$, which originate from the oscillations of the dynamical variables and their conjugate momenta. The appearance of $\omega \sqrt{n}$ in the equations of motion~\eqref{eq:eomphi} provides further support for the observed behavior $p/\tau\sim \omega \sqrt{n_c}$ (see Fig.~\ref{fig:number of bangs}).

\emph{Conclusions.---}As a first application of optimal control to braiding non-Abelian anyons, we obtained bang-bang protocols that can move Majorana zero modes along a quantum wire in \textit{finite} times, while reducing the associated nonadiabatic errors by orders of magnitude (compared with naive smooth protocols). Our calculations were based on a figure of merit that maximizes the magnitude of the overlap between the resulting wave function and the adiabatic one. While more sophisticated cost functions might be needed to account for, e.g., phase errors in a realistic braiding process, our results suggest that optimal control could play an important role in topological quantum computing. Adiabatic braiding can achieve remarkable robustness at the expense of performance. By beating the barrier of adiabaticity, our optimal-control approach may foster the development of high-performance topological quantum computers.

We acknowledge valuable discussions with Chang-Yu Hou  and Falko Pientka. This work was funded by the Packard Foundation and the Institute for Quantum Information and Matter, an NSF Physics Frontiers Center with support of the Gordon and Betty Moore Foundation through Grant GBMF1250 (TK,GR). We also acknowledge support from the U.S. DOE through LANL/LDRD program, NSERC, CIfAR and Max Planck - UBC Centre for Quantum Materials (AR) as well as the Helmholtz Virtual Institute "New states of matter and their excitations" and SPP1285 of the Deutsche Forschungsgemeinschaft (FvO).

\bibliography{controlling.bib}

%merlin.mbs apsrev4-1.bst 2010-07-25 4.21a (PWD, AO, DPC) hacked
%Control: key (0)
%Control: author (8) initials jnrlst
%Control: editor formatted (1) identically to author
%Control: production of article title (-1) disabled
%Control: page (0) single
%Control: year (1) truncated
%Control: production of eprint (0) enabled
\begin{thebibliography}{40}%
\makeatletter
\providecommand \@ifxundefined [1]{%
 \@ifx{#1\undefined}
}%
\providecommand \@ifnum [1]{%
 \ifnum #1\expandafter \@firstoftwo
 \else \expandafter \@secondoftwo
 \fi
}%
\providecommand \@ifx [1]{%
 \ifx #1\expandafter \@firstoftwo
 \else \expandafter \@secondoftwo
 \fi
}%
\providecommand \natexlab [1]{#1}%
\providecommand \enquote  [1]{``#1''}%
\providecommand \bibnamefont  [1]{#1}%
\providecommand \bibfnamefont [1]{#1}%
\providecommand \citenamefont [1]{#1}%
\providecommand \href@noop [0]{\@secondoftwo}%
\providecommand \href [0]{\begingroup \@sanitize@url \@href}%
\providecommand \@href[1]{\@@startlink{#1}\@@href}%
\providecommand \@@href[1]{\endgroup#1\@@endlink}%
\providecommand \@sanitize@url [0]{\catcode `\\12\catcode `\$12\catcode
  `\&12\catcode `\#12\catcode `\^12\catcode `\_12\catcode `\%12\relax}%
\providecommand \@@startlink[1]{}%
\providecommand \@@endlink[0]{}%
\providecommand \url  [0]{\begingroup\@sanitize@url \@url }%
\providecommand \@url [1]{\endgroup\@href {#1}{\urlprefix }}%
\providecommand \urlprefix  [0]{URL }%
\providecommand \Eprint [0]{\href }%
\providecommand \doibase [0]{http://dx.doi.org/}%
\providecommand \selectlanguage [0]{\@gobble}%
\providecommand \bibinfo  [0]{\@secondoftwo}%
\providecommand \bibfield  [0]{\@secondoftwo}%
\providecommand \translation [1]{[#1]}%
\providecommand \BibitemOpen [0]{}%
\providecommand \bibitemStop [0]{}%
\providecommand \bibitemNoStop [0]{.\EOS\space}%
\providecommand \EOS [0]{\spacefactor3000\relax}%
\providecommand \BibitemShut  [1]{\csname bibitem#1\endcsname}%
\let\auto@bib@innerbib\@empty
%</preamble>
\bibitem [{\citenamefont {Kitaev}(2003)}]{kitaev_fault-tolerant_2003}%
  \BibitemOpen
  \bibfield  {author} {\bibinfo {author} {\bibfnamefont {A.}~\bibnamefont
  {Kitaev}},\ }\href {\doibase 10.1016/S0003-4916(02)00018-0} {\bibfield
  {journal} {\bibinfo  {journal} {Ann. Phys.}\ }\textbf {\bibinfo {volume}
  {303}},\ \bibinfo {pages} {2} (\bibinfo {year} {2003})}\BibitemShut {NoStop}%
\bibitem [{\citenamefont {Nayak}\ \emph {et~al.}(2008)\citenamefont {Nayak},
  \citenamefont {Simon}, \citenamefont {Stern}, \citenamefont {Freedman},\ and\
  \citenamefont {Das~Sarma}}]{nayak_non-abelian_2008}%
  \BibitemOpen
  \bibfield  {author} {\bibinfo {author} {\bibfnamefont {C.}~\bibnamefont
  {Nayak}}, \bibinfo {author} {\bibfnamefont {S.~H.}\ \bibnamefont {Simon}},
  \bibinfo {author} {\bibfnamefont {A.}~\bibnamefont {Stern}}, \bibinfo
  {author} {\bibfnamefont {M.}~\bibnamefont {Freedman}}, \ and\ \bibinfo
  {author} {\bibfnamefont {S.}~\bibnamefont {Das~Sarma}},\ }\href {\doibase
  10.1103/RevModPhys.80.1083} {\bibfield  {journal} {\bibinfo  {journal} {Rev.
  Mod. Phys.}\ }\textbf {\bibinfo {volume} {80}},\ \bibinfo {pages} {1083}
  (\bibinfo {year} {2008})}\BibitemShut {NoStop}%
\bibitem [{\citenamefont {Cheng}\ \emph {et~al.}(2011)\citenamefont {Cheng},
  \citenamefont {Galitski},\ and\ \citenamefont
  {Das~Sarma}}]{cheng_nonadiabatic_2011}%
  \BibitemOpen
  \bibfield  {author} {\bibinfo {author} {\bibfnamefont {M.}~\bibnamefont
  {Cheng}}, \bibinfo {author} {\bibfnamefont {V.}~\bibnamefont {Galitski}}, \
  and\ \bibinfo {author} {\bibfnamefont {S.}~\bibnamefont {Das~Sarma}},\ }\href
  {\doibase 10.1103/PhysRevB.84.104529} {\bibfield  {journal} {\bibinfo
  {journal} {Phys. Rev. B}\ }\textbf {\bibinfo {volume} {84}},\ \bibinfo
  {pages} {104529} (\bibinfo {year} {2011})}\BibitemShut {NoStop}%
\bibitem [{\citenamefont {Perfetto}(2013)}]{perfetto_dynamical_2013}%
  \BibitemOpen
  \bibfield  {author} {\bibinfo {author} {\bibfnamefont {E.}~\bibnamefont
  {Perfetto}},\ }\href {\doibase 10.1103/PhysRevLett.110.087001} {\bibfield
  {journal} {\bibinfo  {journal} {Phys. Rev. Lett.}\ }\textbf {\bibinfo
  {volume} {110}},\ \bibinfo {pages} {087001} (\bibinfo {year}
  {2013})}\BibitemShut {NoStop}%
\bibitem [{\citenamefont {Karzig}\ \emph {et~al.}(2013)\citenamefont {Karzig},
  \citenamefont {Refael},\ and\ \citenamefont {von
  Oppen}}]{karzig_boosting_2013}%
  \BibitemOpen
  \bibfield  {author} {\bibinfo {author} {\bibfnamefont {T.}~\bibnamefont
  {Karzig}}, \bibinfo {author} {\bibfnamefont {G.}~\bibnamefont {Refael}}, \
  and\ \bibinfo {author} {\bibfnamefont {F.}~\bibnamefont {von Oppen}},\ }\href
  {\doibase 10.1103/PhysRevX.3.041017} {\bibfield  {journal} {\bibinfo
  {journal} {Phys. Rev. X}\ }\textbf {\bibinfo {volume} {3}},\ \bibinfo {pages}
  {041017} (\bibinfo {year} {2013})}\BibitemShut {NoStop}%
\bibitem [{\citenamefont {Scheurer}\ and\ \citenamefont
  {Shnirman}(2013)}]{scheurer_nonadiabatic_2013}%
  \BibitemOpen
  \bibfield  {author} {\bibinfo {author} {\bibfnamefont {M.~S.}\ \bibnamefont
  {Scheurer}}\ and\ \bibinfo {author} {\bibfnamefont {A.}~\bibnamefont
  {Shnirman}},\ }\href {\doibase 10.1103/PhysRevB.88.064515} {\bibfield
  {journal} {\bibinfo  {journal} {Phys. Rev. B}\ }\textbf {\bibinfo {volume}
  {88}},\ \bibinfo {pages} {064515} (\bibinfo {year} {2013})}\BibitemShut
  {NoStop}%
\bibitem [{\citenamefont {Amorim}\ \emph {et~al.}(2014)\citenamefont {Amorim},
  \citenamefont {Ebihara}, \citenamefont {Yamakage}, \citenamefont {Tanaka},\
  and\ \citenamefont {Sato}}]{amorim_majorana_2014}%
  \BibitemOpen
  \bibfield  {author} {\bibinfo {author} {\bibfnamefont {C.~S.}\ \bibnamefont
  {Amorim}}, \bibinfo {author} {\bibfnamefont {K.}~\bibnamefont {Ebihara}},
  \bibinfo {author} {\bibfnamefont {A.}~\bibnamefont {Yamakage}}, \bibinfo
  {author} {\bibfnamefont {Y.}~\bibnamefont {Tanaka}}, \ and\ \bibinfo {author}
  {\bibfnamefont {M.}~\bibnamefont {Sato}},\ }\href
  {http://arxiv.org/abs/1405.5153} {\bibfield  {journal} {\bibinfo  {journal}
  {{arXiv}:1405.5153}\ } (\bibinfo {year} {2014})}\BibitemShut {NoStop}%
\bibitem [{\citenamefont {Alicea}(2012)}]{alicea_new_2012}%
  \BibitemOpen
  \bibfield  {author} {\bibinfo {author} {\bibfnamefont {J.}~\bibnamefont
  {Alicea}},\ }\href {http://arxiv.org/abs/1202.1293} {\bibfield  {journal}
  {\bibinfo  {journal} {Rep. Prog. Phys.}\ }\textbf {\bibinfo {volume} {75}},\
  \bibinfo {pages} {076501} (\bibinfo {year} {2012})}\BibitemShut {NoStop}%
\bibitem [{\citenamefont {Beenakker}(2013)}]{beenakker_search_2013}%
  \BibitemOpen
  \bibfield  {author} {\bibinfo {author} {\bibfnamefont {C.}~\bibnamefont
  {Beenakker}},\ }\href {\doibase 10.1146/annurev-conmatphys-030212-184337}
  {\bibfield  {journal} {\bibinfo  {journal} {Annu. Rev. Condens. Matter
  Phys.}\ }\textbf {\bibinfo {volume} {4}},\ \bibinfo {pages} {113} (\bibinfo
  {year} {2013})}\BibitemShut {NoStop}%
\bibitem [{\citenamefont {Fu}\ and\ \citenamefont
  {Kane}(2008)}]{fu_superconducting_2008}%
  \BibitemOpen
  \bibfield  {author} {\bibinfo {author} {\bibfnamefont {L.}~\bibnamefont
  {Fu}}\ and\ \bibinfo {author} {\bibfnamefont {C.~L.}\ \bibnamefont {Kane}},\
  }\href {\doibase 10.1103/PhysRevLett.100.096407} {\bibfield  {journal}
  {\bibinfo  {journal} {Phys. Rev. Lett.}\ }\textbf {\bibinfo {volume} {100}},\
  \bibinfo {pages} {096407} (\bibinfo {year} {2008})}\BibitemShut {NoStop}%
\bibitem [{\citenamefont {Lutchyn}\ \emph {et~al.}(2010)\citenamefont
  {Lutchyn}, \citenamefont {Sau},\ and\ \citenamefont
  {Das~Sarma}}]{lutchyn_majorana_2010}%
  \BibitemOpen
  \bibfield  {author} {\bibinfo {author} {\bibfnamefont {R.~M.}\ \bibnamefont
  {Lutchyn}}, \bibinfo {author} {\bibfnamefont {J.~D.}\ \bibnamefont {Sau}}, \
  and\ \bibinfo {author} {\bibfnamefont {S.}~\bibnamefont {Das~Sarma}},\ }\href
  {\doibase 10.1103/PhysRevLett.105.077001} {\bibfield  {journal} {\bibinfo
  {journal} {Phys. Rev. Lett.}\ }\textbf {\bibinfo {volume} {105}},\ \bibinfo
  {pages} {077001} (\bibinfo {year} {2010})}\BibitemShut {NoStop}%
\bibitem [{\citenamefont {Oreg}\ \emph {et~al.}(2010)\citenamefont {Oreg},
  \citenamefont {Refael},\ and\ \citenamefont {von Oppen}}]{oreg_helical_2010}%
  \BibitemOpen
  \bibfield  {author} {\bibinfo {author} {\bibfnamefont {Y.}~\bibnamefont
  {Oreg}}, \bibinfo {author} {\bibfnamefont {G.}~\bibnamefont {Refael}}, \ and\
  \bibinfo {author} {\bibfnamefont {F.}~\bibnamefont {von Oppen}},\ }\href
  {\doibase 10.1103/PhysRevLett.105.177002} {\bibfield  {journal} {\bibinfo
  {journal} {Phys. Rev. Lett.}\ }\textbf {\bibinfo {volume} {105}},\ \bibinfo
  {pages} {177002} (\bibinfo {year} {2010})}\BibitemShut {NoStop}%
\bibitem [{\citenamefont {Duckheim}\ and\ \citenamefont
  {Brouwer}(2011)}]{duckheim_andreev_2011}%
  \BibitemOpen
  \bibfield  {author} {\bibinfo {author} {\bibfnamefont {M.}~\bibnamefont
  {Duckheim}}\ and\ \bibinfo {author} {\bibfnamefont {P.~W.}\ \bibnamefont
  {Brouwer}},\ }\href {\doibase 10.1103/PhysRevB.83.054513} {\bibfield
  {journal} {\bibinfo  {journal} {Phys. Rev. B}\ }\textbf {\bibinfo {volume}
  {83}},\ \bibinfo {pages} {054513} (\bibinfo {year} {2011})}\BibitemShut
  {NoStop}%
\bibitem [{\citenamefont {Nadj-Perge}\ \emph {et~al.}(2013)\citenamefont
  {Nadj-Perge}, \citenamefont {Drozdov}, \citenamefont {Bernevig},\ and\
  \citenamefont {Yazdani}}]{nadj-perge_proposal_2013}%
  \BibitemOpen
  \bibfield  {author} {\bibinfo {author} {\bibfnamefont {S.}~\bibnamefont
  {Nadj-Perge}}, \bibinfo {author} {\bibfnamefont {I.~K.}\ \bibnamefont
  {Drozdov}}, \bibinfo {author} {\bibfnamefont {B.~A.}\ \bibnamefont
  {Bernevig}}, \ and\ \bibinfo {author} {\bibfnamefont {A.}~\bibnamefont
  {Yazdani}},\ }\href {\doibase 10.1103/PhysRevB.88.020407} {\bibfield
  {journal} {\bibinfo  {journal} {Phys. Rev. B}\ }\textbf {\bibinfo {volume}
  {88}},\ \bibinfo {pages} {020407} (\bibinfo {year} {2013})}\BibitemShut
  {NoStop}%
\bibitem [{\citenamefont {Mourik}\ \emph {et~al.}(2012)\citenamefont {Mourik},
  \citenamefont {Zuo}, \citenamefont {Frolov}, \citenamefont {Plissard},
  \citenamefont {Bakkers},\ and\ \citenamefont
  {Kouwenhoven}}]{mourik_signatures_2012}%
  \BibitemOpen
  \bibfield  {author} {\bibinfo {author} {\bibfnamefont {V.}~\bibnamefont
  {Mourik}}, \bibinfo {author} {\bibfnamefont {K.}~\bibnamefont {Zuo}},
  \bibinfo {author} {\bibfnamefont {S.~M.}\ \bibnamefont {Frolov}}, \bibinfo
  {author} {\bibfnamefont {S.~R.}\ \bibnamefont {Plissard}}, \bibinfo {author}
  {\bibfnamefont {E.~P. a.~M.}\ \bibnamefont {Bakkers}}, \ and\ \bibinfo
  {author} {\bibfnamefont {L.~P.}\ \bibnamefont {Kouwenhoven}},\ }\href
  {\doibase 10.1126/science.1222360} {\bibfield  {journal} {\bibinfo  {journal}
  {Science}\ }\textbf {\bibinfo {volume} {336}},\ \bibinfo {pages} {1003}
  (\bibinfo {year} {2012})}\BibitemShut {NoStop}%
\bibitem [{\citenamefont {Das}\ \emph {et~al.}(2012)\citenamefont {Das},
  \citenamefont {Ronen}, \citenamefont {Most}, \citenamefont {Oreg},
  \citenamefont {Heiblum},\ and\ \citenamefont
  {Shtrikman}}]{das_zero-bias_2012}%
  \BibitemOpen
  \bibfield  {author} {\bibinfo {author} {\bibfnamefont {A.}~\bibnamefont
  {Das}}, \bibinfo {author} {\bibfnamefont {Y.}~\bibnamefont {Ronen}}, \bibinfo
  {author} {\bibfnamefont {Y.}~\bibnamefont {Most}}, \bibinfo {author}
  {\bibfnamefont {Y.}~\bibnamefont {Oreg}}, \bibinfo {author} {\bibfnamefont
  {M.}~\bibnamefont {Heiblum}}, \ and\ \bibinfo {author} {\bibfnamefont
  {H.}~\bibnamefont {Shtrikman}},\ }\href {\doibase 10.1038/nphys2479}
  {\bibfield  {journal} {\bibinfo  {journal} {Nat. Phys.}\ }\textbf {\bibinfo
  {volume} {8}},\ \bibinfo {pages} {887} (\bibinfo {year} {2012})}\BibitemShut
  {NoStop}%
\bibitem [{\citenamefont {Deng}\ \emph {et~al.}(2012)\citenamefont {Deng},
  \citenamefont {Yu}, \citenamefont {Huang}, \citenamefont {Larsson},
  \citenamefont {Caroff},\ and\ \citenamefont {Xu}}]{deng_anomalous_2012}%
  \BibitemOpen
  \bibfield  {author} {\bibinfo {author} {\bibfnamefont {M.~T.}\ \bibnamefont
  {Deng}}, \bibinfo {author} {\bibfnamefont {C.~L.}\ \bibnamefont {Yu}},
  \bibinfo {author} {\bibfnamefont {G.~Y.}\ \bibnamefont {Huang}}, \bibinfo
  {author} {\bibfnamefont {M.}~\bibnamefont {Larsson}}, \bibinfo {author}
  {\bibfnamefont {P.}~\bibnamefont {Caroff}}, \ and\ \bibinfo {author}
  {\bibfnamefont {H.~Q.}\ \bibnamefont {Xu}},\ }\href {\doibase
  10.1021/nl303758w} {\bibfield  {journal} {\bibinfo  {journal} {Nano Lett.}\
  }\textbf {\bibinfo {volume} {12}},\ \bibinfo {pages} {6414} (\bibinfo {year}
  {2012})}\BibitemShut {NoStop}%
\bibitem [{\citenamefont {Rokhinson}\ \emph {et~al.}(2012)\citenamefont
  {Rokhinson}, \citenamefont {Liu},\ and\ \citenamefont
  {Furdyna}}]{rokhinson_fractional_2012}%
  \BibitemOpen
  \bibfield  {author} {\bibinfo {author} {\bibfnamefont {L.~P.}\ \bibnamefont
  {Rokhinson}}, \bibinfo {author} {\bibfnamefont {X.}~\bibnamefont {Liu}}, \
  and\ \bibinfo {author} {\bibfnamefont {J.~K.}\ \bibnamefont {Furdyna}},\
  }\href {\doibase 10.1038/nphys2429} {\bibfield  {journal} {\bibinfo
  {journal} {Nat Phys}\ }\textbf {\bibinfo {volume} {8}},\ \bibinfo {pages}
  {795} (\bibinfo {year} {2012})}\BibitemShut {NoStop}%
\bibitem [{\citenamefont {Finck}\ \emph {et~al.}(2013)\citenamefont {Finck},
  \citenamefont {Van~Harlingen}, \citenamefont {Mohseni}, \citenamefont
  {Jung},\ and\ \citenamefont {Li}}]{finck_anomalous_2013}%
  \BibitemOpen
  \bibfield  {author} {\bibinfo {author} {\bibfnamefont {A.~D.~K.}\
  \bibnamefont {Finck}}, \bibinfo {author} {\bibfnamefont {D.~J.}\ \bibnamefont
  {Van~Harlingen}}, \bibinfo {author} {\bibfnamefont {P.~K.}\ \bibnamefont
  {Mohseni}}, \bibinfo {author} {\bibfnamefont {K.}~\bibnamefont {Jung}}, \
  and\ \bibinfo {author} {\bibfnamefont {X.}~\bibnamefont {Li}},\ }\href
  {\doibase 10.1103/PhysRevLett.110.126406} {\bibfield  {journal} {\bibinfo
  {journal} {Phys. Rev. Lett.}\ }\textbf {\bibinfo {volume} {110}},\ \bibinfo
  {pages} {126406} (\bibinfo {year} {2013})}\BibitemShut {NoStop}%
\bibitem [{\citenamefont {Pribiag}\ \emph {et~al.}(2014)\citenamefont
  {Pribiag}, \citenamefont {Beukman}, \citenamefont {Qu}, \citenamefont
  {Cassidy}, \citenamefont {Charpentier}, \citenamefont {Wegscheider},\ and\
  \citenamefont {Kouwenhoven}}]{pribiag_edge-mode_2014}%
  \BibitemOpen
  \bibfield  {author} {\bibinfo {author} {\bibfnamefont {V.~S.}\ \bibnamefont
  {Pribiag}}, \bibinfo {author} {\bibfnamefont {A.~J.~A.}\ \bibnamefont
  {Beukman}}, \bibinfo {author} {\bibfnamefont {F.}~\bibnamefont {Qu}},
  \bibinfo {author} {\bibfnamefont {M.~C.}\ \bibnamefont {Cassidy}}, \bibinfo
  {author} {\bibfnamefont {C.}~\bibnamefont {Charpentier}}, \bibinfo {author}
  {\bibfnamefont {W.}~\bibnamefont {Wegscheider}}, \ and\ \bibinfo {author}
  {\bibfnamefont {L.~P.}\ \bibnamefont {Kouwenhoven}},\ }\href
  {http://arxiv.org/abs/1408.1701} {\bibfield  {journal} {\bibinfo  {journal}
  {{arXiv}:1408.1701}\ } (\bibinfo {year} {2014})}\BibitemShut {NoStop}%
\bibitem [{\citenamefont {Hart}\ \emph {et~al.}(2014)\citenamefont {Hart},
  \citenamefont {Ren}, \citenamefont {Wagner}, \citenamefont {Leubner},
  \citenamefont {M{\"u}hlbauer}, \citenamefont {Br{\"u}ne}, \citenamefont
  {Buhmann}, \citenamefont {Molenkamp},\ and\ \citenamefont
  {Yacoby}}]{hart_induced_2014}%
  \BibitemOpen
  \bibfield  {author} {\bibinfo {author} {\bibfnamefont {S.}~\bibnamefont
  {Hart}}, \bibinfo {author} {\bibfnamefont {H.}~\bibnamefont {Ren}}, \bibinfo
  {author} {\bibfnamefont {T.}~\bibnamefont {Wagner}}, \bibinfo {author}
  {\bibfnamefont {P.}~\bibnamefont {Leubner}}, \bibinfo {author} {\bibfnamefont
  {M.}~\bibnamefont {M{\"u}hlbauer}}, \bibinfo {author} {\bibfnamefont
  {C.}~\bibnamefont {Br{\"u}ne}}, \bibinfo {author} {\bibfnamefont
  {H.}~\bibnamefont {Buhmann}}, \bibinfo {author} {\bibfnamefont {L.~W.}\
  \bibnamefont {Molenkamp}}, \ and\ \bibinfo {author} {\bibfnamefont
  {A.}~\bibnamefont {Yacoby}},\ }\href {\doibase 10.1038/nphys3036} {\bibfield
  {journal} {\bibinfo  {journal} {Nat Phys}\ }\textbf {\bibinfo {volume}
  {10}},\ \bibinfo {pages} {638} (\bibinfo {year} {2014})}\BibitemShut
  {NoStop}%
\bibitem [{\citenamefont {Nadj-Perge}\ \emph {et~al.}(2014)\citenamefont
  {Nadj-Perge}, \citenamefont {Drozdov}, \citenamefont {Li}, \citenamefont
  {Chen}, \citenamefont {Jeon}, \citenamefont {Seo}, \citenamefont {MacDonald},
  \citenamefont {Bernevig},\ and\ \citenamefont
  {Yazdani}}]{nadj-perge_observation_2014}%
  \BibitemOpen
  \bibfield  {author} {\bibinfo {author} {\bibfnamefont {S.}~\bibnamefont
  {Nadj-Perge}}, \bibinfo {author} {\bibfnamefont {I.~K.}\ \bibnamefont
  {Drozdov}}, \bibinfo {author} {\bibfnamefont {J.}~\bibnamefont {Li}},
  \bibinfo {author} {\bibfnamefont {H.}~\bibnamefont {Chen}}, \bibinfo {author}
  {\bibfnamefont {S.}~\bibnamefont {Jeon}}, \bibinfo {author} {\bibfnamefont
  {J.}~\bibnamefont {Seo}}, \bibinfo {author} {\bibfnamefont {A.~H.}\
  \bibnamefont {MacDonald}}, \bibinfo {author} {\bibfnamefont {B.~A.}\
  \bibnamefont {Bernevig}}, \ and\ \bibinfo {author} {\bibfnamefont
  {A.}~\bibnamefont {Yazdani}},\ }\href {\doibase 10.1126/science.1259327}
  {\bibfield  {journal} {\bibinfo  {journal} {Science}\ }\textbf {\bibinfo
  {volume} {346}},\ \bibinfo {pages} {602} (\bibinfo {year}
  {2014})}\BibitemShut {NoStop}%
\bibitem [{\citenamefont {Alicea}\ \emph {et~al.}(2011)\citenamefont {Alicea},
  \citenamefont {Oreg}, \citenamefont {Refael}, \citenamefont {von Oppen},\
  and\ \citenamefont {Fisher}}]{alicea_non-abelian_2011}%
  \BibitemOpen
  \bibfield  {author} {\bibinfo {author} {\bibfnamefont {J.}~\bibnamefont
  {Alicea}}, \bibinfo {author} {\bibfnamefont {Y.}~\bibnamefont {Oreg}},
  \bibinfo {author} {\bibfnamefont {G.}~\bibnamefont {Refael}}, \bibinfo
  {author} {\bibfnamefont {F.}~\bibnamefont {von Oppen}}, \ and\ \bibinfo
  {author} {\bibfnamefont {M.~P.~A.}\ \bibnamefont {Fisher}},\ }\href {\doibase
  10.1038/nphys1915} {\bibfield  {journal} {\bibinfo  {journal} {Nat. Phys.}\
  }\textbf {\bibinfo {volume} {7}},\ \bibinfo {pages} {412} (\bibinfo {year}
  {2011})}\BibitemShut {NoStop}%
\bibitem [{\citenamefont {Goldstein}\ and\ \citenamefont
  {Chamon}(2011)}]{goldstein_decay_2011}%
  \BibitemOpen
  \bibfield  {author} {\bibinfo {author} {\bibfnamefont {G.}~\bibnamefont
  {Goldstein}}\ and\ \bibinfo {author} {\bibfnamefont {C.}~\bibnamefont
  {Chamon}},\ }\href {\doibase 10.1103/PhysRevB.84.205109} {\bibfield
  {journal} {\bibinfo  {journal} {Phys. Rev. B}\ }\textbf {\bibinfo {volume}
  {84}},\ \bibinfo {pages} {205109} (\bibinfo {year} {2011})}\BibitemShut
  {NoStop}%
\bibitem [{\citenamefont {Budich}\ \emph {et~al.}(2012)\citenamefont {Budich},
  \citenamefont {Walter},\ and\ \citenamefont
  {Trauzettel}}]{budich_failure_2012}%
  \BibitemOpen
  \bibfield  {author} {\bibinfo {author} {\bibfnamefont {J.~C.}\ \bibnamefont
  {Budich}}, \bibinfo {author} {\bibfnamefont {S.}~\bibnamefont {Walter}}, \
  and\ \bibinfo {author} {\bibfnamefont {B.}~\bibnamefont {Trauzettel}},\
  }\href {\doibase 10.1103/PhysRevB.85.121405} {\bibfield  {journal} {\bibinfo
  {journal} {Phys. Rev. B}\ }\textbf {\bibinfo {volume} {85}},\ \bibinfo
  {pages} {121405} (\bibinfo {year} {2012})}\BibitemShut {NoStop}%
\bibitem [{\citenamefont {Rainis}\ and\ \citenamefont
  {Loss}(2012)}]{rainis_majorana_2012}%
  \BibitemOpen
  \bibfield  {author} {\bibinfo {author} {\bibfnamefont {D.}~\bibnamefont
  {Rainis}}\ and\ \bibinfo {author} {\bibfnamefont {D.}~\bibnamefont {Loss}},\
  }\href {\doibase 10.1103/PhysRevB.85.174533} {\bibfield  {journal} {\bibinfo
  {journal} {Phys. Rev. B}\ }\textbf {\bibinfo {volume} {85}},\ \bibinfo
  {pages} {174533} (\bibinfo {year} {2012})}\BibitemShut {NoStop}%
\bibitem [{\citenamefont {Hohenester}\ \emph {et~al.}(2007)\citenamefont
  {Hohenester}, \citenamefont {Rekdal}, \citenamefont {Borz{\`i}},\ and\
  \citenamefont {Schmiedmayer}}]{hohenester_optimal_2007}%
  \BibitemOpen
  \bibfield  {author} {\bibinfo {author} {\bibfnamefont {U.}~\bibnamefont
  {Hohenester}}, \bibinfo {author} {\bibfnamefont {P.~K.}\ \bibnamefont
  {Rekdal}}, \bibinfo {author} {\bibfnamefont {A.}~\bibnamefont {Borz{\`i}}}, \
  and\ \bibinfo {author} {\bibfnamefont {J.}~\bibnamefont {Schmiedmayer}},\
  }\href {\doibase 10.1103/PhysRevA.75.023602} {\bibfield  {journal} {\bibinfo
  {journal} {Phys. Rev. A}\ }\textbf {\bibinfo {volume} {75}},\ \bibinfo
  {pages} {023602} (\bibinfo {year} {2007})}\BibitemShut {NoStop}%
\bibitem [{\citenamefont {Salamon}\ \emph {et~al.}(2009)\citenamefont
  {Salamon}, \citenamefont {Hoffmann}, \citenamefont {Rezek},\ and\
  \citenamefont {Kosloff}}]{salamon_maximum_2009}%
  \BibitemOpen
  \bibfield  {author} {\bibinfo {author} {\bibfnamefont {P.}~\bibnamefont
  {Salamon}}, \bibinfo {author} {\bibfnamefont {K.~H.}\ \bibnamefont
  {Hoffmann}}, \bibinfo {author} {\bibfnamefont {Y.}~\bibnamefont {Rezek}}, \
  and\ \bibinfo {author} {\bibfnamefont {R.}~\bibnamefont {Kosloff}},\ }\href
  {\doibase 10.1039/B816102J} {\bibfield  {journal} {\bibinfo  {journal} {Phys.
  Chem. Chem. Phys.}\ }\textbf {\bibinfo {volume} {11}},\ \bibinfo {pages}
  {1027} (\bibinfo {year} {2009})}\BibitemShut {NoStop}%
\bibitem [{\citenamefont {Chen}\ \emph {et~al.}(2010)\citenamefont {Chen},
  \citenamefont {Ruschhaupt}, \citenamefont {Schmidt}, \citenamefont {del
  Campo}, \citenamefont {Gu{\'e}ry-Odelin},\ and\ \citenamefont
  {Muga}}]{chen_fast_2010}%
  \BibitemOpen
  \bibfield  {author} {\bibinfo {author} {\bibfnamefont {X.}~\bibnamefont
  {Chen}}, \bibinfo {author} {\bibfnamefont {A.}~\bibnamefont {Ruschhaupt}},
  \bibinfo {author} {\bibfnamefont {S.}~\bibnamefont {Schmidt}}, \bibinfo
  {author} {\bibfnamefont {A.}~\bibnamefont {del Campo}}, \bibinfo {author}
  {\bibfnamefont {D.}~\bibnamefont {Gu{\'e}ry-Odelin}}, \ and\ \bibinfo
  {author} {\bibfnamefont {J.~G.}\ \bibnamefont {Muga}},\ }\href {\doibase
  10.1103/PhysRevLett.104.063002} {\bibfield  {journal} {\bibinfo  {journal}
  {Phys. Rev. Lett.}\ }\textbf {\bibinfo {volume} {104}},\ \bibinfo {pages}
  {063002} (\bibinfo {year} {2010})}\BibitemShut {NoStop}%
\bibitem [{\citenamefont {Rahmani}\ and\ \citenamefont
  {Chamon}(2011)}]{rahmani_optimal_2011}%
  \BibitemOpen
  \bibfield  {author} {\bibinfo {author} {\bibfnamefont {A.}~\bibnamefont
  {Rahmani}}\ and\ \bibinfo {author} {\bibfnamefont {C.}~\bibnamefont
  {Chamon}},\ }\href {\doibase 10.1103/PhysRevLett.107.016402} {\bibfield
  {journal} {\bibinfo  {journal} {Phys. Rev. Lett.}\ }\textbf {\bibinfo
  {volume} {107}},\ \bibinfo {pages} {016402} (\bibinfo {year}
  {2011})}\BibitemShut {NoStop}%
\bibitem [{\citenamefont {Doria}\ \emph {et~al.}(2011)\citenamefont {Doria},
  \citenamefont {Calarco},\ and\ \citenamefont
  {Montangero}}]{doria_optimal_2011}%
  \BibitemOpen
  \bibfield  {author} {\bibinfo {author} {\bibfnamefont {P.}~\bibnamefont
  {Doria}}, \bibinfo {author} {\bibfnamefont {T.}~\bibnamefont {Calarco}}, \
  and\ \bibinfo {author} {\bibfnamefont {S.}~\bibnamefont {Montangero}},\
  }\href {\doibase 10.1103/PhysRevLett.106.190501} {\bibfield  {journal}
  {\bibinfo  {journal} {Phys. Rev. Lett.}\ }\textbf {\bibinfo {volume} {106}},\
  \bibinfo {pages} {190501} (\bibinfo {year} {2011})}\BibitemShut {NoStop}%
\bibitem [{\citenamefont {Caneva}\ \emph {et~al.}(2011)\citenamefont {Caneva},
  \citenamefont {Calarco}, \citenamefont {Fazio}, \citenamefont {Santoro},\
  and\ \citenamefont {Montangero}}]{caneva_speeding_2011}%
  \BibitemOpen
  \bibfield  {author} {\bibinfo {author} {\bibfnamefont {T.}~\bibnamefont
  {Caneva}}, \bibinfo {author} {\bibfnamefont {T.}~\bibnamefont {Calarco}},
  \bibinfo {author} {\bibfnamefont {R.}~\bibnamefont {Fazio}}, \bibinfo
  {author} {\bibfnamefont {G.~E.}\ \bibnamefont {Santoro}}, \ and\ \bibinfo
  {author} {\bibfnamefont {S.}~\bibnamefont {Montangero}},\ }\href {\doibase
  10.1103/PhysRevA.84.012312} {\bibfield  {journal} {\bibinfo  {journal} {Phys.
  Rev. A}\ }\textbf {\bibinfo {volume} {84}},\ \bibinfo {pages} {012312}
  (\bibinfo {year} {2011})}\BibitemShut {NoStop}%
\bibitem [{\citenamefont {Choi}\ \emph {et~al.}(2011)\citenamefont {Choi},
  \citenamefont {Onofrio},\ and\ \citenamefont
  {Sundaram}}]{choi_optimized_2011}%
  \BibitemOpen
  \bibfield  {author} {\bibinfo {author} {\bibfnamefont {S.}~\bibnamefont
  {Choi}}, \bibinfo {author} {\bibfnamefont {R.}~\bibnamefont {Onofrio}}, \
  and\ \bibinfo {author} {\bibfnamefont {B.}~\bibnamefont {Sundaram}},\ }\href
  {\doibase 10.1103/PhysRevA.84.051601} {\bibfield  {journal} {\bibinfo
  {journal} {Phys. Rev. A}\ }\textbf {\bibinfo {volume} {84}},\ \bibinfo
  {pages} {051601} (\bibinfo {year} {2011})}\BibitemShut {NoStop}%
\bibitem [{\citenamefont {del Campo}(2011)}]{del_campo_fast_2011}%
  \BibitemOpen
  \bibfield  {author} {\bibinfo {author} {\bibfnamefont {A.}~\bibnamefont {del
  Campo}},\ }\href {\doibase 10.1209/0295-5075/96/60005} {\bibfield  {journal}
  {\bibinfo  {journal} {{EPL} (Europhysics Letters)}\ }\textbf {\bibinfo
  {volume} {96}},\ \bibinfo {pages} {60005} (\bibinfo {year}
  {2011})}\BibitemShut {NoStop}%
\bibitem [{\citenamefont {Hoffmann}\ \emph {et~al.}(2011)\citenamefont
  {Hoffmann}, \citenamefont {Salamon}, \citenamefont {Rezek},\ and\
  \citenamefont {Kosloff}}]{hoffmann_time-optimal_2011}%
  \BibitemOpen
  \bibfield  {author} {\bibinfo {author} {\bibfnamefont {K.~H.}\ \bibnamefont
  {Hoffmann}}, \bibinfo {author} {\bibfnamefont {P.}~\bibnamefont {Salamon}},
  \bibinfo {author} {\bibfnamefont {Y.}~\bibnamefont {Rezek}}, \ and\ \bibinfo
  {author} {\bibfnamefont {R.}~\bibnamefont {Kosloff}},\ }\href {\doibase
  10.1209/0295-5075/96/60015} {\bibfield  {journal} {\bibinfo  {journal}
  {{EPL}}\ }\textbf {\bibinfo {volume} {96}},\ \bibinfo {pages} {60015}
  (\bibinfo {year} {2011})}\BibitemShut {NoStop}%
\bibitem [{\citenamefont {Rahmani}\ \emph {et~al.}(2013)\citenamefont
  {Rahmani}, \citenamefont {Kitagawa}, \citenamefont {Demler},\ and\
  \citenamefont {Chamon}}]{rahmani_cooling_2013}%
  \BibitemOpen
  \bibfield  {author} {\bibinfo {author} {\bibfnamefont {A.}~\bibnamefont
  {Rahmani}}, \bibinfo {author} {\bibfnamefont {T.}~\bibnamefont {Kitagawa}},
  \bibinfo {author} {\bibfnamefont {E.}~\bibnamefont {Demler}}, \ and\ \bibinfo
  {author} {\bibfnamefont {C.}~\bibnamefont {Chamon}},\ }\href {\doibase
  10.1103/PhysRevA.87.043607} {\bibfield  {journal} {\bibinfo  {journal} {Phys.
  Rev. A}\ }\textbf {\bibinfo {volume} {87}},\ \bibinfo {pages} {043607}
  (\bibinfo {year} {2013})}\BibitemShut {NoStop}%
\bibitem [{Note1()}]{Note1}%
  \BibitemOpen
  \bibinfo {note} {With the exception {$\delimiter "426830A \phi ^{(0)}_n|\phi
  ^{(v)}_{-n}\delimiter "526930B \propto (v/u)^{2}$}. See also Supplemental
  Information.}\BibitemShut {Stop}%
\bibitem [{\citenamefont {Rahmani}(2013)}]{rahmani_quantum_2013}%
  \BibitemOpen
  \bibfield  {author} {\bibinfo {author} {\bibfnamefont {A.}~\bibnamefont
  {Rahmani}},\ }\href {\doibase 10.1142/S0217984913300196} {\bibfield
  {journal} {\bibinfo  {journal} {Mod. Phys. Lett. B}\ }\textbf {\bibinfo
  {volume} {27}},\ \bibinfo {pages} {1330019} (\bibinfo {year}
  {2013})}\BibitemShut {NoStop}%
\bibitem [{\citenamefont {Akhmerov}(2010)}]{akhmerov_topological_2010}%
  \BibitemOpen
  \bibfield  {author} {\bibinfo {author} {\bibfnamefont {A.~R.}\ \bibnamefont
  {Akhmerov}},\ }\href {\doibase 10.1103/PhysRevB.82.020509} {\bibfield
  {journal} {\bibinfo  {journal} {Phys. Rev. B}\ }\textbf {\bibinfo {volume}
  {82}},\ \bibinfo {pages} {020509} (\bibinfo {year} {2010})}\BibitemShut
  {NoStop}%
\bibitem [{\citenamefont {Pontryagin}(1987)}]{pontryagin_mathematical_1987}%
  \BibitemOpen
  \bibfield  {author} {\bibinfo {author} {\bibfnamefont {L.~S.}\ \bibnamefont
  {Pontryagin}},\ }\href@noop {} {\emph {\bibinfo {title} {Mathematical Theory
  of Optimal Processes}}}\ (\bibinfo  {publisher} {{CRC} Press},\ \bibinfo
  {year} {1987})\BibitemShut {NoStop}%
\end{thebibliography}%

\beginsupplement

\clearpage

\widetext

\begin{center}
	\textbf{\large Supplemental Information}
\end{center}
\section{S\arabic{section}: Finite velocity wavefunctions}
\addtocounter{section}{1}

The finite velocity bound state wavefunctions $\phi_n^{(v_i)}[x-y(t)]$ can be obtained by applying a Lorentz boost to Eq.~(2) and take the form [5]
\begin{eqnarray}
\phi_{n}^{(v)}(x) & = & \gamma^{1/4}\left(\!\!\begin{array}{cc}
\sqrt{1+v/u} & 0\\
0 & \sqrt{1-v/u}
\end{array}\right)\phi_{n}\left(\sqrt{\gamma}x\right){\rm e}^{{\rm i}\sqrt{2\gamma n}(v/u)(x/\xi)}\,,
\label{eq:phi}
\end{eqnarray}
where $\gamma=1/\sqrt{1-(v/u)^2}$ and the renormalized bound state spectrum is given by $\varepsilon_{n}^{(v)}={\rm sign}(n)\gamma^{-3/2}\sqrt{|n|}\omega$. The crucial difference of the finite-velocity, relative to the static bound states is the momentum boost of the form $\exp({\rm i}q x)$. Since the $\phi_n$ consist of harmonic oscillator wavefunctions $g_n$ the corresponding matrix elements $\langle \phi_k^{(v)} | \phi_n\rangle$ are controlled by integrals
\begin{equation}
\int{\rm d}xg_{n}(x)g_{n'}(x){\rm e}^{-{\rm i}\sqrt{2}qx/\xi} =
 {\rm e}^{-\frac{1}{2}q^{2}}\sqrt{m!/M!}\left(-{\rm i}q\right)^{M-m}\!\! L_{m}^{M-m}\left(q^{2}\right),
 \label{eq:Laguerre}
\end{equation}
where $M=\max(n,n')$, $m=\min(n,n')$, and $L_m^{M-m}$ are associated Laguerre polynomials. Since $L_m^{M-m}(q=0)$ just contributes with a constant one obtains by applying Eq.~(\ref{eq:Laguerre}) that 
\begin{equation}
\langle \phi_k^{(v)} | \phi_n\rangle\propto (v/u)^{||k|-|n||}
\label{eq:overlap}
\end{equation}
 to leading order in $v/u$, with the exception of the $k=-n$ term that is proportional to $(v/u)^2$. Specifically, to linear order in $v/u$, we only obtain nonvanishing contributions 
\begin{equation}
\langle \phi_k^{(v)} | \phi_n\rangle = \frac{1}{4} (-{\rm i}M v/u)\left( 1 + {\rm sign}(k n)\sqrt{(|M|-1)/|M|}\right)\ \ \ ,\textrm{if }||k|-|n||=1\,,
\label{eq:overlap2}
\end{equation}
and $\langle \phi_k^{(v)} | \phi_n\rangle = 1+{\rm sign}(k n)$, if $|k|=|n|$. Here, $M$ is again given by the larger (in absolute value) of $k$ and $n$.

\section{S\arabic{section}: Evaluation of the cost function}
\addtocounter{section}{1}

To evaluate the cost function $c(\tau)$ it is helpful to express Eq.~(3) explicitly in terms of the Heisenberg operators $\hat{\gamma}_i(\tau)=U(\tau)^\dagger\hat{\gamma}_{n,B}U(\tau)$,
\begin{equation}
c(T)=\sum_{i\geq 0_+}\left\langle\hat{\gamma}_{i,B}^\dagger(\tau) \hat{\gamma}_{i,B}(\tau) \right\rangle_0 - \sum_{j>i;\,i\geq 0_+} \left\langle \hat{\gamma}_{j,B}^\dagger(\tau) \hat{\gamma}_{i,B}^\dagger(\tau) \hat{\gamma}_{i,B}(\tau) \hat{\gamma}_{j,B}(\tau)\right\rangle_0+\dots\,,
\label{eq:normalorder}
\end{equation}
where all expectation values $\langle \dots \rangle_0$ are taken with respect to the initial ground state with all states at $i\geq 0_+$ unoccupied. Note that we use the shorthand notation $\hat{\gamma}_{0_+,B} \equiv \hat{d}_0=(\hat{\gamma}_{0,B}+{\rm i}\hat{\gamma}_{0,C})/\sqrt{2}$ and $\hat{\gamma}_{0_-,B} \equiv \hat{d}_0^\dagger$. From the normal ordered form of Eq.~(\ref{eq:normalorder}) it becomes clear that nonvanishing contributions to the cost function $c(\tau)$ require transitions of initial $\hat{\gamma}_{i\geq 0_+,B}$ to final $\hat{\gamma}_{j\geq 0_+,A}^\dagger$ during the Heisenberg evolution (as mentioned in the main text). With the knowledge of the time evolution [see Eq.~(4)] we can write $\hat{\gamma}_{i,B}(\tau)=\sum_j \alpha_{ij}\hat{\gamma}_{{j,A}}$, where the sum over $j$ runs from $-n_{\rm max}$, over $0$, to $+n_{\rm max}$. Note that there is a subtlety in treating the zero modes. To transform from the basis using the Majorana operators $\hat{\gamma}_{0,B}$ to the fermionic zero modes $\hat{\gamma}_{0_{\pm},B}$ used in Eq.~(\ref{eq:normalorder}) we define the $(2n_{\rm max}+2)\times (2n_{\rm max}+2)$ dimensional matrix $\tilde{\alpha}_{nm}$ such that
\begin{equation}
\left(\begin{array}{c}
	\hat{\gamma}_{n>0,B}(\tau)\\
	\hat{d}_{0}(\tau)\\
	\hat{d}^{\dagger}_{0}(\tau)\\
	\hat{\gamma}_{n<0,B}(\tau)
\end{array}\right) = \sum_m
\underbrace{
	\left(\begin{array}{cccc}
		\alpha_{nm} & \frac{1}{\sqrt{2}}\alpha_{n,0} & \frac{1}{\sqrt{2}}\alpha_{n,0} & \alpha_{nm}\\
		\frac{1}{\sqrt{2}}\alpha_{0,m} & \frac{\alpha_{00}+1}{2} & \frac{\alpha_{00}-1}{2} & \frac{1}{\sqrt{2}}\alpha_{0,m}\\
		\frac{1}{\sqrt{2}}\alpha_{0,m} & \frac{\alpha_{00}-1}{2} & \frac{\alpha_{00}+1}{2} & \frac{1}{\sqrt{2}}\alpha_{0,m}\\
		\alpha_{nm} & \frac{1}{\sqrt{2}}\alpha_{n,0} & \frac{1}{\sqrt{2}}\alpha_{n,0} & \alpha_{nm}
	\end{array}\right)
}_{\tilde{\alpha}_{nm}}\left(\begin{array}{c}
\hat{\gamma}_{m>0,A}\\
\hat{d}_0\\
\hat{d}_0^{\dagger}\\
\hat{\gamma}_{m<0,A}
\end{array}\right)\,.
\label{eq:alpha2alpha}
\end{equation}
This allows to express the cost function as
\begin{equation}
c(\tau)=\sum_{i,j\geq 0_+} \tilde{\alpha}_{-i,j} \tilde{\alpha}_{i,-j} - \sum_{i>j;\,i,j,k,l\geq 0_+} \left( \tilde{\alpha}_{-j,k} \tilde{\alpha}_{-i,-k} \tilde{\alpha}_{i,l} \tilde{\alpha}_{j,-l} + \tilde{\alpha}_{-j,k} \tilde{\alpha}_{-i,l} \tilde{\alpha}_{i,-l} \tilde{\alpha}_{j,-k} - \tilde{\alpha}_{-j,k} \tilde{\alpha}_{-i,l} \tilde{\alpha}_{i,-k} \tilde{\alpha}_{j,-l} \right) + \dots
\label{eq:costalpha}
\end{equation}
As mentioned in the main text the expansion of Eq.~(\ref{eq:costalpha}) is ultimately controlled by the velocity of the domain wall $v$. For small $v/u$ the matrix $\tilde{\alpha}_{ij}$ is mainly diagonal as the off-diagonal terms are suppressed by powers of $(v/u)$ [see Eq.~(\ref{eq:overlap})]. One can therefore obtain an estimate of the importance of the different terms in Eq.~(\ref{eq:costalpha}) by counting orders of $v/u$. The first term involves two off-diagonal elements and is therefore of order $(v/u)^2$. Interestingly, the second term is of the same order since for $k=i=l$ it takes the same form and is only smaller by a factor of two than the first term because of the restriction of the sum to $i>j$. The third and fourth terms are already of order $(v/u)^4$. By writing higher order terms, e.g., $\langle \delta n_i \delta n_j \delta n_k \rangle$ in normal ordered form similar to  Eq.~(\ref{eq:normalorder}), one can quickly show that they are also suppressed by at least $(v/u)^4$ which justifies  Eq.~(3).

\begin{figure}
\includegraphics[scale=0.55]{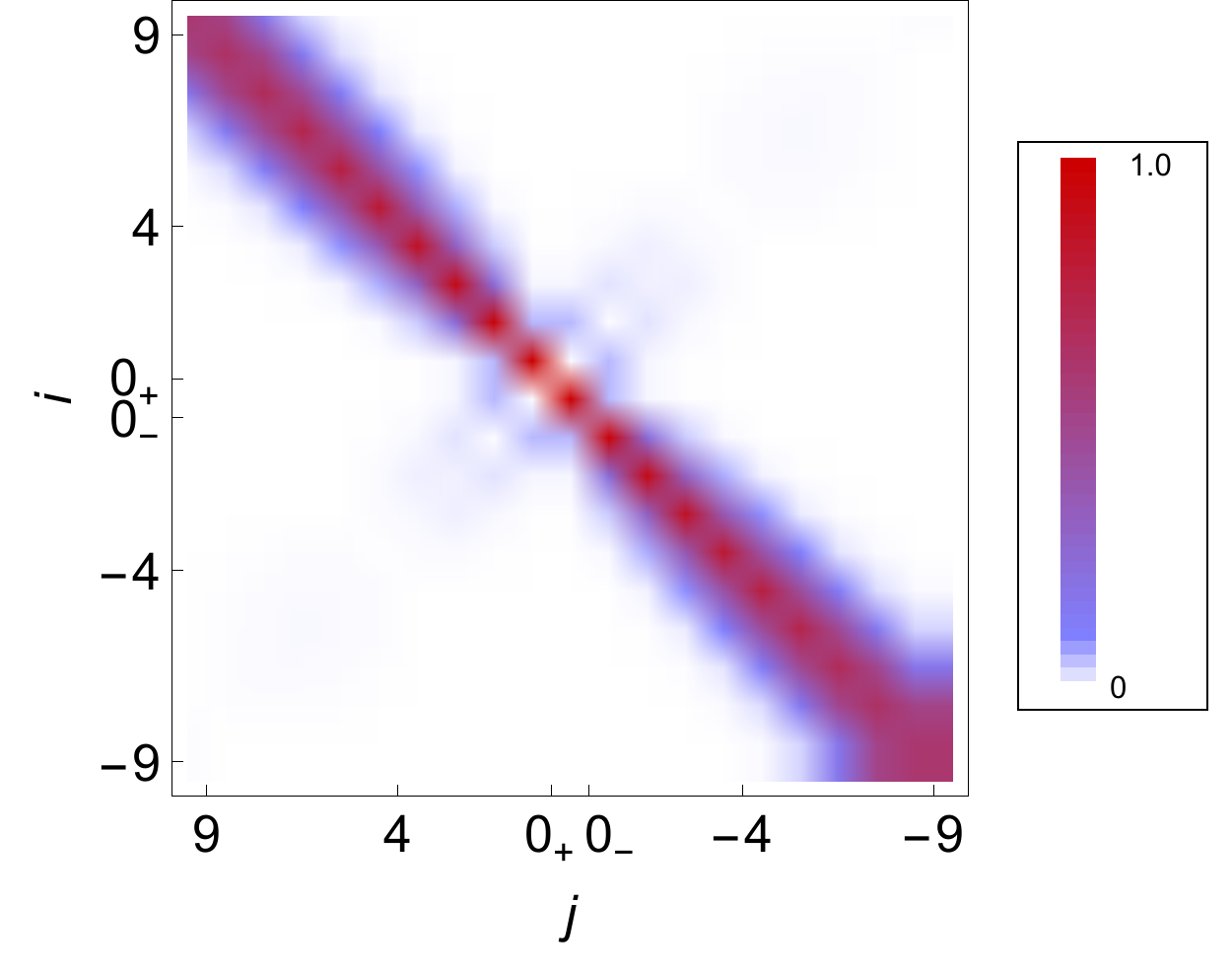}
\caption{Array plot of the matrix $|\tilde{\alpha}_{ij}|$ describing the Heisenberg time evolution as $\hat{\gamma}_{i,B}(\tau)=\sum_j\tilde{\alpha}_{ij}\hat{\gamma}_{j,A}$. The corresponding protocol is an optimal bang-bang protocol depicted in Fig.~3 with $\tau=3/\omega$, $v_{\rm max}=0.3u$, $n_{\rm max}=30$, and $n_c=7$.\label{fig:alpha}}
\end{figure}

Note that although the small $v/u$ limit gives a convenient way to quantify the above expansion, the approximation remains well justified even for moderate $v/u$ as long as the time evolution does not create too many excitations. In fact moderate $v/u$ allow values of $\tilde{\alpha}_{i,j}$ to be of order 1 even for $i\neq j$ as long as $i$ and $j$ both have the same sign. However, due to the Pauli principle, these processes cannot cause changes in the occupation numbers which require transitions from negative to positive energy states. The latter are still rare for not too non-adiabatic evolutions as can be seen from the suppression of the off-diagonal blocks in Fig.~\ref{fig:alpha} (see also Eq.~(\ref{eq:overlap2}). On can then use the number of occurrences of $\tilde{\alpha}_{i,j}$ with $\mathrm{sign}(ij)=-1$ in Eq.~(\ref{eq:costalpha}) to replace the small parameter $v/u$, which yields the same terms in the expansion.

\section{S\arabic{section}: Occupation of the high energy states}
\addtocounter{section}{1}

\begin{figure}[center]
	\includegraphics[scale=0.9]{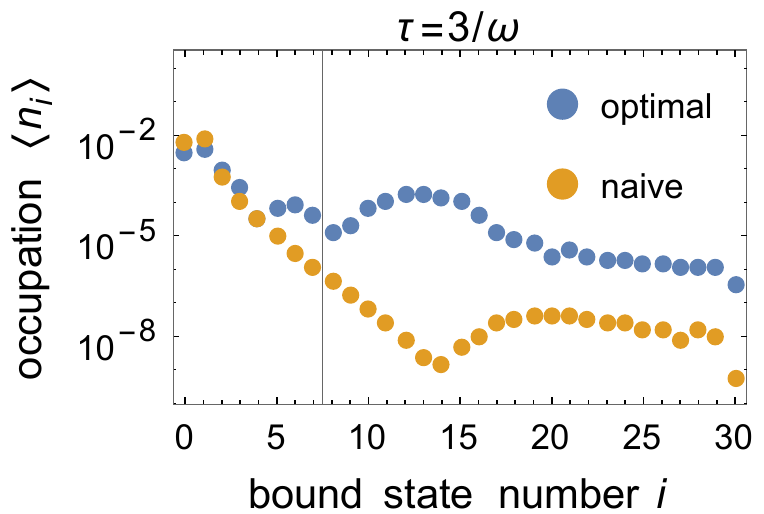}\hspace{1cm}\includegraphics[scale=.9]{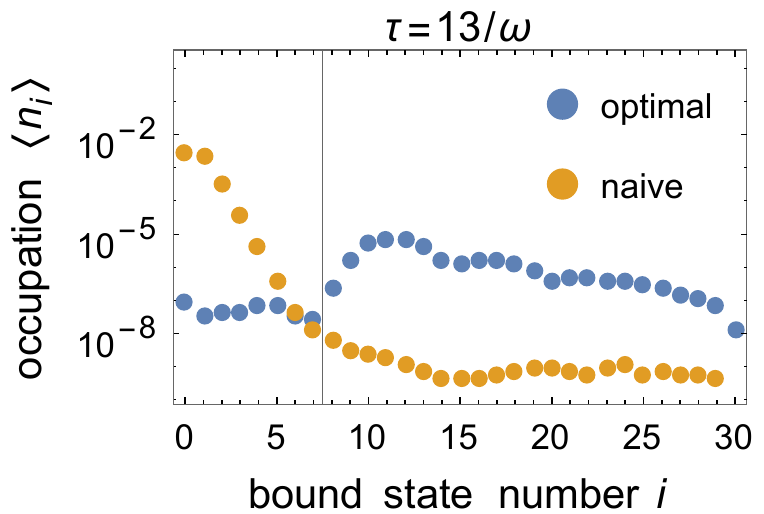}
	\caption{Comparison of the bound state occupation $\langle n_i \rangle$ of optimal and naive Gaussian protocols. The vertical line indicates the separation between the (low energy) states included in the cost function ($i\leq n_c=7$) and the (high energy) states left out of the optimization ($i>n_c$). The left and right panel show the behavior for short and long protocol durations $\tau$. Other parameters are $v_{\rm max}=0.3u$ and $n_{\rm max}=30$.}
	\label{fig:high_energy}
\end{figure}

The optimization minimizes the nonadiabatic occupations of the $n_c+1$ (positive energy) bound states included in the cost function (referred to as low energy states in the following). The occupation of higher energy ($i>n_c$) and possible continuum states, however, is not necessarily lower for the optimal protocol when compared to naive protocols. In fact, in most cases the low energy states are optimized to the expense of an increased number of high energy excitations. Figure~\ref{fig:high_energy} illustrates this effect for in the limit of short ($\tau=3/\omega$) and long protocol durations ($\tau=13/\omega$). For short durations the optimization only improves the most problematic lowest energy states, while the general trend of decreasing nonadiabatic occupations with increasing energy remains. For long protocol durations the optimization leads to a strong suppression of the occupation of all low energy states (close to the numerical precision). In this regime, the vanishing occupation of the low energy states of the optimal protocol makes the high energy states (not to far away from energies $\varepsilon_{n_c}$) the leading excitations. Note, however, that the high energy excitations are still much weaker than the uncorrected low energy excitations of the naive protocol. The latter effect follows from the natural suppression of high energy excitations for not-to-stongly diabatic protocols.

In addition to the weak occupation of the high energy states their effect on the Majorana modes is also limited by other constraints. For a unitary time evolution within the non-interacting electronic degrees of freedom considered here, a finite excitation of high energy degrees of freedom has no effect on the delocalized fermionic mode $\hat{d}_0$ (and its parity) that carries the quantum information of the system. More specifically, the Heisenberg time evolution reads (in the notation of section S2)
\begin{equation}
U(\tau)^\dagger\hat{d}_0 U(\tau)=\frac{1}{\sqrt{2}}\left( \sum_j\alpha_{0,j}\hat{\gamma}_{j,A}+\mathrm{i}\hat{\gamma}_{0,C}\right) \ \overset{\mathrm{optimization}}{\longrightarrow}\  \hat{d}_0\,.
\label{eq:elastic_evolution}
\end{equation}
The crucial observation is that since $\hat{\gamma}_{0,C}$ is static it will not be affected by the time evolution even when including the excitation of higher energy and possible continuum states. The optimization of the moving low energy states ensures that $\alpha_{0,j}=\delta_{0,j}$ is fulfilled as closely as possible.

Equation~\eqref{eq:elastic_evolution} describes the absence of inelastic processes, where the high energy decrees of freedom are decoupled from the (static) lowest energy bound states. Interestingly, the quantum information is also protected in the opposite limit of strong inelastic processes (for example caused by coupling to a low temperature bosonic bath). When relaxation effects are strong the excited high energy states recombine quickly before they can propagate to the far-away Majorana mode $\hat{\gamma}_{0,C}$.

\section{S\arabic{section}: Effect of the bound state number $n_c$}
\addtocounter{section}{1}
 
\begin{figure}[center]
	\includegraphics[scale=0.9]{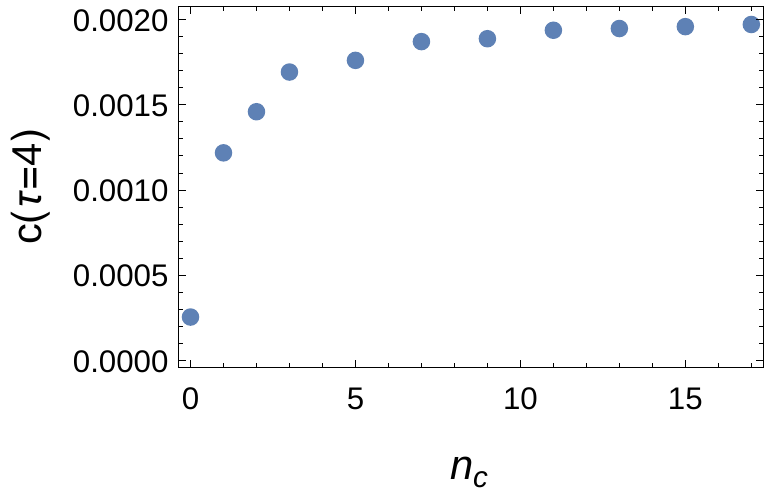}\hspace{1cm}\includegraphics[scale=.9]{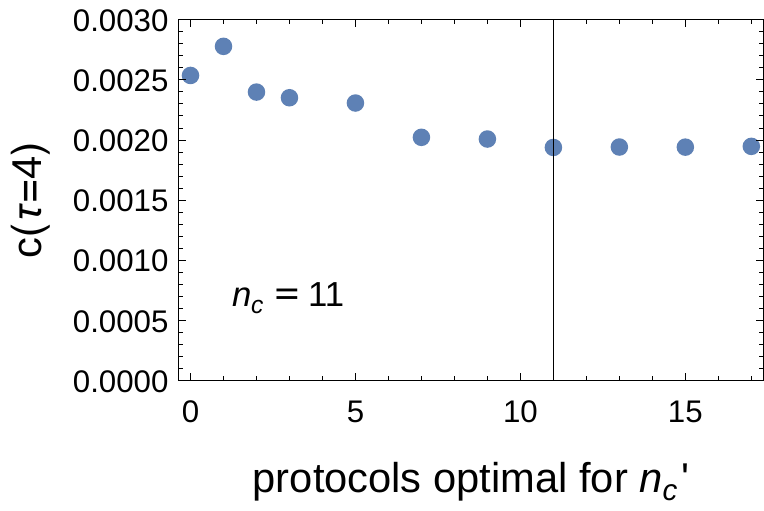}
	\caption{Dependence of protocol performance on $n_c$ for the example $\tau=4/\omega$. Left panel: Change of the cost function with the included number of bound states $n_c$. Including more states increases the cost function, nevertheless, $c(\tau)$ seems to converge for large $n_c$. Right panel: Applying the cost function with $n_c=11$ to protocols optimized for  $n_c'$. Although the protocol optimized for $n_c'=n_c$ performs best, the differences to other protocols (showing different $p$) becomes minuscule in the regime of large $n_c,n_c'$.}
	\label{fig:cost_vs_nc}
\end{figure}

As mentioned in the main text the number of high velocity plateaus $p$ increases when taking into account an increasing number of bound states $n_c$ in the optimization scheme. The lack of protocol convergence for $n_c \rightarrow \infty$ might seem surprising at first but is not problematic because of the decreasing importance of the protocol change. The left panel of Fig.~\ref{fig:cost_vs_nc} suggests that the cost function saturates in the $n_c \rightarrow \infty$ limit. Moreover, when measuring the protocol performance with respect to a number of bound states $n_c$, it might even be sufficient to optimize the protocols for a different $n_c'$ and still obtain similarly good results. This is shown in the right panel of Fig.~\ref{fig:cost_vs_nc}. Both of these observations reflect the fact that the leading error that contributes to the cost function is caused by excitations of low energy states (roughly $n_c \lesssim 5$ in this case). Once the optimization addresses these leading contributions the cost function starts saturating.

\section{S\arabic{section}: Connection with Pontryagin's maximum principle}

As discussed in the main text, the bang-bang nature of the protocols can be understood in terms of Pontryagin's theorem. Since we have a large number of dynamical variables ${\rm Re} (\varphi^m_n)$, ${\rm Re} (\theta^m_n)$, ${\rm Im} (\varphi^m_n)$, ${\rm Im} (\theta^m_n)$, a direct solution of the Pontryagin equations to obtain the optimal protocol is difficult. However, once we have a protocol from Monte Carlo simulations, consistency with the Pontryagin equations provides a valuable check. The equations of motion given in the main text for the dynamical variables can be derived from the Schr\"odinger equation $-i\partial_t\psi^m(x,t)={\cal H}\psi^m(x,t)$ [see Eq. (2) of the main text], which leads to
\begin{equation}\label{eq:eom1}
\sum_{n=0}^\infty
\left(\begin{array}{c}
\dot{\varphi}_n\\ 
\dot{\theta}_n
\end{array} \right)
g_n(\tilde{x})
=
\sum_{n=0}^\infty\bigg[
\left(\begin{array}{c}
\varphi_n\\ 
\theta_n
\end{array} \right)
\dot{y}(t)
g'_n(\tilde{x})
-
u\left(\begin{array}{c}
\varphi_n\\ 
-\theta_n
\end{array} \right)
g'_n(\tilde{x})
+
ib \tilde{x}\left(\begin{array}{c}
\theta_n\\ 
\varphi_n
\end{array} \right)
g_n(\tilde{x})
\bigg],
\end{equation}
where we have suppressed the superscript $m$ for the dynamical variables and $\tilde{x}\equiv x-y(t)$. We now use the following properties of the Hermite polynomials $H'_n(z)=2 n H_{n-1}(z)$ and $zH_n(z)=nH_{n-1}(z)+{1\over 2} H_{n+1}(z)$
to write
\begin{equation}
 g'_n(z)={1\over \xi}\left[\sqrt{n\over 2} g_{n-1}(z)-\sqrt{n+1\over 2} g_{n+1}(z)\right],\qquad
z g_n(z)={ \xi}\left[\sqrt{n\over 2} g_{n-1}(z)+\sqrt{n+1\over 2} g_{n+1}(z)\right],
\end{equation}
which upon insertion into Eq.~\eqref{eq:eom1}, and using $\int_{-\infty}^{+\infty} g_n(x)g_m(x)dx=\delta_{nm}$, leads to the equations of motion in Eq. (7) of the main text. To evaluate the cost function at the end of the time evolution we expand the single-particle wave function $\psi^m(x,\tau)$ in terms of the domain-wall bound-state wave functions at position $B$ as $\psi^m(x,\tau)=\sum_n \alpha_{m,n}^* \phi_n(x-\ell)$, where
\begin{equation}
\alpha_{m,n}^*=(\varphi^m_n-i\theta^m_n-i\varphi^m_{n-1}+\theta^m_{n-1})/2,\qquad \alpha_{m,-n}^*=(\varphi^m_n-i\theta^m_n+i\varphi^m_{n-1}-\theta^m_{n-1})/2, \qquad n>0.
\end{equation}
With the knowledge of $\alpha_{m,n}$ the cost function can then be evaluated using Eqs.~(\ref{eq:alpha2alpha}) and (\ref{eq:costalpha}).

\begin{figure}[center]
	\includegraphics[scale=0.2]{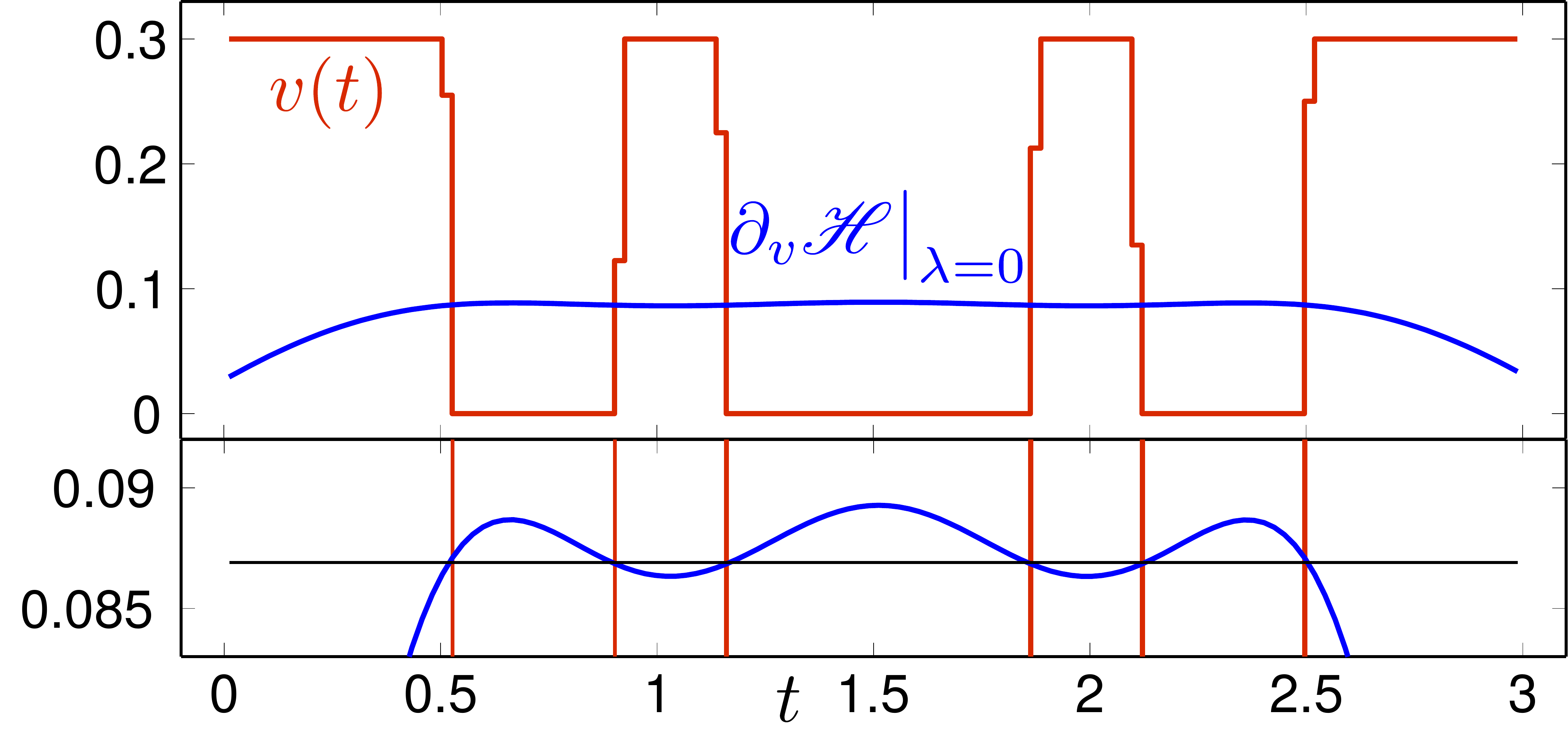}
	\caption{Upper panel: An optimal protocol $v(t)$ for $\tau=3$, $v_{\rm max}=0.3$, $v_{\rm ave}=0.15$, $n_c=7$ (red line). The coefficient of $v$ in the optimal-control Hamiltonian $\mathscr H$ with the unknown Lagrange multiplier $\lambda$ set to zero (blue curve). Lower panel: A zoom-in of the same plot. With a shift corresponding to $\lambda$ the protocol can be determined by ${\rm sgn}(\partial_v {\mathscr H})$.}
	\label{fig:pontry}
\end{figure}

To verify the Pontryagin equations, we introduce conjugate momenta $\Pi_{{\rm Re}(\varphi^m_n)}$, $\Pi_{{\rm Re} (\theta^m_n)}$, $\Pi_{{\rm Im} (\varphi^m_n)}$, $\Pi_{{\rm Im} (\theta^m_n)}$ for each of the dynamical variables, write the optimal control Hamiltonian, and derive the equations of motion for the conjugate momenta (which turn out to be very similar to the equations of motions for the dynamical variables). Given the protocol, we find the dynamical variables as a functions of time. From the derivative of the cost function with respect to the dynamical variables, we obtain the final values of the conjugate momenta and evaluate them for all times by solving their equations of motion backward in time. We can then calculate $\partial_v {\mathscr H}$ up to the unknown Lagrange multiplier $\lambda$, which simply shifts this quantity. The consistency of our numerical protocols with Pontryagin's equations requires $\partial_v {\mathscr H}=0$ at the jumps in the protocol (because this allows values of $v(t)$ different from $0$ or $v_{\rm max}$). Without the knowledge of the shift by $\lambda$, this simply implies that  $\partial_v {\mathscr H}$ must take the same values for \textit{all} times that coincide with the jumps in the protocol, which is precisely what we obtain as shown in Fig.~\ref{fig:pontry}. 

\end{document}